\documentstyle[worldsci,epsfig]{article}
\input epsf
\newcommand{\be}{\begin{eqnarray}}
\newcommand{\ee}{\end{eqnarray}}

\begin{document}

  \title {
Scales and Phases 
   of  Non-Perturbative QCD
  }
  \author {
     E.V.~Shuryak\\
     Department of Physics and Astronomy,\\ State University of New York, 
     Stony Brook, NY 11794-3800 USA
  }
  \date{\today}
  \maketitle    

\begin{abstract}
This is a short write-up of lectures at UK Theory Institute, Swansea UK, Sept.99; and XVII School
``QCD: Perturbative or Non-perturbative?'' Lisbon,Portugal, Oct.99.
The covered topic include (i) discussion of the scales at which
perturbative description should be changed by the non-perturbative
one. A number of various examples are considered, leading to different
scales,
 related to
instantons or confinement effects; (ii) Instanton
vacuum as an Instanton Liquid; (iii) The phase diagram of QCD at finite
temperature
and density, and especially description of
a recent progress in Color Superconductivity. 

\end{abstract}

\vspace{0.1in}

\section{Introduction}
  I hardly need  an explanation of why I have devoted so large
portion of these lectures to such introductory subject as ``scales"
of non-perturbative QCD. First of all, it answers the question put in
the
title of the Lisbon school. The second reason:
somewhat surprisingly to me, naive simplistic ideas originating in
70's - the picture hadrons as some structureless perturbative 
``bags", with very soft non-perturbative effects appearing only
at its boundaries at the scale 1 fm --
 are still alive and well, in spite of multiple evidences to
the contrary, pointed to existence of some substructures
inside hadrons. The lectures in such schools
is  probably the right place  to collect at least few simplest
 evidences for that in one place.

  Indeed, the perturbative QCD naturally points toward the $\Lambda_{QCD}\sim
200 \ MeV$, the position of the so called Landau pole,  as 
the momentum scale where it becomes inapplicable.
Since the inverse of it, 1 fm, is a typical hadronic size,
 such cutoff indeed seemed natural. The basic approach
to non-perturbative physics which has started in
70's, the QCD sum rules, have been based on such a picture.
If the non-perturbative fields be so ``soft", 
the derivative expansion,
known as Operator Product Expansion (OPE),
 should be reasonable convergent. Furthermore, 
 smooth connection between hadronic and partonic approaches
hoped to be possible.

  However it was realized in 80's  
  that
the non-perturbative fields actually form structures
with sizes significantly smaller than 1 fm. 
Two most important non-perturbative objects
are instantons (with the average size $\rho \sim 1/3 fm$ \cite{Shu_82})
and confining strings (with the radius of the gluoelectric field
only about $r_{string}\approx 1/5  fm$, see fig.2b ). 
the ratio of those to hadronic size R enters in the
4-th power for $\rho$  and in the 2-th for  $r_{string}$,
so in both cases only a small fraction, few percent, of hadronic
volume is occupied by them. 
The simplest picture of it is that of constituent quarks (or, possibly,
scalar diquarks as well) of the size $\sim\rho$, connected by strings.

Can we see it in some observables? The simplest thing is to monitor
the onset of the non-perturbative effects at small distances/
large momenta where these 
phenomena  just become comparable to perturbative effects,
and, furthermore, the OPE expansion in derivatives start to fail. 
 Where it happens would define the scales we speak about, as you see
the number
depends on a particular physical problem considered:
several different situation which
 have been identified so far will be discussed in this section.
For pedagogical reason, we do it in two rounds: at the first one
(section 2) we
present the ``naive" predictions and, in many cases, see if those
really work. 

In the second round (the next sections 3,4) we discuss the explanations
 (to the extent they are understood today).  
  The conclusions we draw from these observations
 are that in several observables already   the distances  
$r=0.1-0.2$ used in these studies are already {\em large enough} to be outside
the validity domain of 
 the original quark-gluon description, even reinforced by the 
 OPE corrections.
Effective approaches   like Interacting Instanton Liquid Model
(IILM) or Abelian Higgs Model (AHM) should
 rather be used here, and those indeed  provide
at least semi-quantitative explanations of
the non-perturbative effects.

One of the reasons the ``naive" point of view mentioned above has
survived for so long is that many applications deal with a special case
of vector currents. Those are related with 
such long-studied reactions as the $e^+e^-\rightarrow$ hadrons
and deep inelastic scattering\footnote{Including neutrino scattering:
the leading twist operators providing moments of the structure
functions are again vectors. The reason they correspond to the amplitude squared. Only for polarized structure functions another channel - the axial current - appears, and that lead to a surprise known as the ``spin crisis".}.
The QCD sum rules work well for this cases, e.g. the ``next twist"
correction to deep inelastic scattering are surprisingly small, etc.
(We will show some of this in section 5, when we will discuss vector correlators.)
However,   if one looks more systematically at various channels (see e.g.
review \cite{Shu_93}), it is clear that the behavior of the vector channels
 is in fact a single exception. Cancellation of all corrections
in this case is observed but not yet fully understood. It seems like
vector currents simply cannot see these vacuum/hadronic substructure
 at all.

We go into more detailed discussion of the instanton-induced
effects in chapter 4. It is shown there that consistent treatment
of those, to all orders in 't Hooft interaction, has been worked out.
The so called Interacting Instanton Liquid Model (IILM) reproduces
all correlation functions known from
phenomenology (hadronic masses and coupling constants) and/or direct
lattice measurements.

We continue to discuss instanton-induced effects in the last chapter 5,
but now in connection with ``new frontiers" in QCD leading to 
qualitatively new phases. Among those are Quark-Gluon Plasma (QGP)
at $high$ T and high baryonic density; Color Superconductivity (CSC)
 at $low$ T and 
high baryonic density, and the Conformal Phase at larger number of flavors.

\section{The Scales of Non-Perturbative QCD}

\subsection{The Chiral Scale} 

Historically the first is the so called
{\em chiral scale}
$\Lambda_\chi\sim 1 \ GeV$, the  upper limit of
 low energy effective
theories
such as effective chiral Lagrangians or
Nambu-Jona-Lasinio model \cite{NJL}. Although it does not follow
that it is also the low boundary of perturbative treatment, it cannot
at least be lower. 

These effective theories appeared in 60's, before QCD was even invented.
Their parameters were simply fitted to observations: e.g. the first
Weinberg term of the chiral Lagrangian contains
2 derivatives and the pion decay constant $f_\pi$,
the next have 4 derivatives and the so called Leutwyler couplings,
 etc. The ``chiral scale" appears when one asks at which momenta
all terms of the chiral Lagrangian become of the same magnitude.

The NJL model is somewhat more microscopic: it does not start with
the Goldstone modes (pions), but derives instead
the chiral symmetry breaking and 
Goldstone modes starting from some hypothetical
 4-fermion  interaction.
I would not describe this model here, and only make few comments about it.
(i) It was the first bridge between theory of superconductivity and 
quantum field theory: in a way, it first shown that the vacuum of the
of strong interaction  should be truly non-perturbative. It has
a non-zero quark condensate and a gap 330-400 MeV, known as ``constituent
quark mass"; (ii) The model has basically 2 parameters: the strength
of its 4-fermion interaction G and the cutoff $\Lambda\sim .8-1 GeV$.
The latter regulates the loops: this model is non-renormalizable.
The latter parameter can also be seen as the inverse size of the constituent
quarks; (iii) The interaction on which NJL model was based is hypothetical,
and we should ask now if QCD actually generates it. People have been first
trying to explain the  NJL forces as being due to one gluon exchange diagrams.
However it became clear by now that it is not so, and the true origin
of the NJL-type short-range quark-antiquark attractions comes from
the instantons; (iv) The symmetries of the resulting 4-fermion effective
interaction, known as 't Hooft Lagrangian, are not the same as of the
original NJL model. They explicitly break this
chiral U(1) symmetry (rotating all fermions by $exp(i\phi*\gamma_5)$), as a result of which
the $SU(3)_f$ singlet meson $\eta'$ is $not$ a Goldstone boson. NJL model
was wrong at this point.

  Instead, we rather consider 
another incarnation of the chiral scale,  defined as the 
energy/momentum scale at which the
non-perturbative effects become equal to perturbative
ones (in fact, the zeroth order in $\alpha_s$) in  $J^{P}=O^{\pm}$ 
correlation functions.

 Let us consider two currents separated by space-like distance x
(the spatial distance, or Euclidean time) and consider the  correlation functions of the type
\be 
K(x)=<T(J(x) J(0))> 
\ee
with $J(x)=\bar \psi(x)  \Gamma \psi(x)$. The matrix
$\Gamma$ contains $\gamma_5$ for pseudoscalar channels, and a flavor matrices,
if we discuss e.g. pions. For simplicity, we will restrict ourselves
in what follows to 2 quark flavors. It means we have 4 channels: $\pi$ (P=-1,
I=1), $\sigma$ (P=+1,I=0), $\eta$ (P=-1,I=0) and 
$\delta$ (or ???) (P=+1,I=1). 

In all cases at small x we expect
$K(x)\approx K_0(x)$ where the latter corresponds to
just free propagation of light quarks.
If they are massless, the correlators are $K_0(x)=12/(\pi^4 x^{6})$,
basically the square of the massless
quark propagator. 

The first deviations due to non-perturbative effects have been calculated
using the OPE  \cite{SVZ},
in the context of
QCD sum rules. 
Ignoring subtleties, one can view it as just expansion in x at small x.
For scalar and pseudoscalar channels
the result for the first correction is
\be
{K(x)\over K_0(x)}=1+ {x^4\over 384} <(gG)^2>+...
\ee
If the ``gluon condensate"  is  made out of soft
 vacuum filed, they are are long range as compared to $x$
and therefore all their arguments can be simply taken at the point x=0.
 The value of the ``gluon condensate" appearing here
was estimated previously from charmonium sum rules:
\be
<(gG)^2>_{SVZ}\approx .5 \, GeV^4
\ee
Thus, the OPE suggests the  scale at which the correction becomes
equal to the first term:
\be
x_{OPE}=(384/<(gG)^2>_{SVZ})^{1/4}\approx 1.0 \, fm
\ee
which seems to be completely consistent with the  expectations.

  However, as the Novikov, Shifman, Vainshtein and Zakharov
soon noticed  \cite{NVSZ}, this (and other OPE corrections) had
completely failed to describe all  $J^{P}=O^{\pm}$ channels.
It was very puzzling, because for vector and axial channels it did
a good job \cite{SVZ}. That is why they asked an important question
(the title
of their paper): ``Are all hadrons alike?".

How do we know that this scale is wrong? Let us just do the simplest
thing: consider the  pion contribution 
to the correlator in question
\be
K_\pi(x) = {\lambda_\pi^2\over 4 \pi^2 x^2} 
\ee
The coupling constant is defined as $ \lambda_\pi=<0|J(0)|\pi>$ and the
rest is nothing else as the scalar massless propagator. (We can ignore
the pion mass at distances in question.)

Because both the pion term and the gluon condensate correction
are $1/x^2$, let us compare the coefficients. Ideal matching
would mean they are about the same\footnote{In fact there are states other the pion,
and all of the contribute positively to the correlator. So one
should rather expect the pion contribution be somewhat smaller than the
correction term, to be corrected by the contribution of other states.}:
\be
\lambda_\pi^2 \approx {<(gG)^2>_{SVZ} \over 8 \pi^2}
\ee
The r.h.s. is about 0.0063 GeV$^4$. However 
 (unlike the better known coupling to
the axial current $f_\pi$), the $ \lambda_\pi$ is surprisingly
large\footnote{The reason for that is the the pion is rather compact and also
the shape of the wave function is concentrates at its center, so that 
its value at r=0 is large. We return to this point in the discussion of
the ``instanton liquid" model.}. The l.h.s. of this relation is actually
 $ \lambda_\pi^2=(.48\, GeV)^4=0.053\, GeV^4$, or about 10 times larger
than the r.h.s. It means some other and much larger effect
should explain deviation from the perturbative behavior.
Why it has not showed up in the vector channels? We return to these
questions in the next section.

\subsection{The Scale of the  $J^{P}=O^{\pm}$ glueball correlation functions} 

It is defined in the same way as the first one, only for scalar and pseudoscalar
glueball correlators. Instead of the quark currents we have now operators
$G_{\mu\nu}^2$ and $G_{\mu\nu}\tilde G_{\mu\nu}$ (tilde means dual
field in electric-magnetic sense, or
convolution with $\epsilon_{\alpha\beta\gamma\delta}/2$.
The analogous OPE expressions for these two
channels \cite{NVSZ}  now starts with the
next gluonic ``condensate" 
\be
K_{J=\pm 1}(x)=\pm{384\over \pi^4 x^8} [1 \pm  {\pi^2 x^6\over 192} <g^3 f^{abc} G_{\mu\nu}^a G_{\nu\sigma}^b G_{\sigma\nu}^c>+...] 
\ee
Again, the first term is the free propagation of gluons, 
 it is the largest one at small x. Note that the power of x
in both terms simply follows from dimension.
  Demanding that these two terms
are equal, we get the corresponding ``OPE scale".

 However, as emphasized in this paper  \cite{NVSZ},
other theoretical arguments suggested that in fact the non-perturbative physics
should start at much smaller scale. 
One of them is based on low energy theorem\footnote{The analogous integral for the pseudoscalar correlator is the so called
topological susceptibility.} 
\be
\int d^4x K_+(x)={128\pi^2 \over b} <(gG)^2>
\ee
where b=${11 N_c\over 3}-{2N_f\over 3}\approx 9$
  is the first coefficient of the beta function.

One way to see what it means is to assume that the major deviation
at small distances is due to the contribution of the {\em scalar glueball}
\footnote{There are evidences that the scalar glueball is
even more compact and stronger bound object than the pion, so it is a reasonable
assumption.}. Then the integral
can be related to  the coupling constant $\lambda_{0+}=<0|G^2| {0+} \ glueball>$ and the glueball mass $m_{0+}$ in a simple case
\be
{\lambda_{0+}^2 \over m_{0+}^2}={128\pi^2 \over b} <(gG)^2>
\ee
It is known from lattice and phenomenology that $m_{0+}\approx 1.6 GeV$,
so we now can obtain the coupling\footnote{Lattice and instanton liquid
calculation of the coupling leads to a number consistent with this estimate
and the low energy theorem is of course exactly satisfied.
}. Then we can go back to  $K_+(x)$ at small x
and see where the glueball contribution becomes equal to the
perturbative part, to define the scale.

Like in the pion case, both the OPE correction and the coupling
produce $1/x^2$ effects at small distances, so we again
compare the coefficients. The OPE-hadron matching would work if they
are close
\be 
\lambda_{0+}^2 \approx 8 <g^3 f^{abc} G_{\mu\nu}^a G_{\nu\sigma}^b G_{\sigma\nu}^c>
\ee
but in fact the value we infer from the low energy theorem gives 1.8 GeV$^6$
for the l.h.s., which again significantly exceeds the 
l.h.s. (when the phenomenological value of the
VEV of $G^3$ operator is substituted.).

More strict way \cite{NVSZ} to say the same thing
is to go to
 momentum representation (Fourier transform $\tilde K_+(q)$
 of the correlation function.  The total integral
over x corresponds to the zero momentum value. Since we know it, we can
subtract it in  the dispersion relation
\be 
\tilde K_+(Q)={1\over \pi}\int ds {Im\tilde K_+(s) \over (s+Q^2)}
\ee
where $Q^2=-q^2>0$. The subtraction
 improves the convergence of the integral 
at large Q of the dispersion relation, namely
\be
{\tilde K_+(Q)\over Q^2} ={\tilde K_+(0)\over Q^2}+ {1 \over \pi}\int ds {Im\tilde K_+(s) \over s(s+Q^2)}
\ee

This argument shows that in momentum space the effective subtraction
term becomes equal to perturbative one at the momentum scale
$\Lambda_{0^{\pm}glueballs} \sim (3-4)  \ GeV$, the
 largest of the effective scales
of the non-perturbative QCD\footnote{
Let me emphasize it once again: it is $not$ the glueball mass scale, but rather the so called gluon-glueball duality
energy. Above it the spectral density can be approximated by the partonic
one, below it cannot.} 
Other set of arguments presented in the same paper have shown,
that the same scale should appear in the pseudoscalar glueball channel
as well. Why this happens
 in this case at  much larger scale than for any quark channels?

\subsection{Heavy Quarkonia and Small Size Instantons as Dynamical Dipoles}

  Another way of looking at the onset of non-perturbative physics
is to separate color charges by some small distance, and see
what happens. (The correlation functions considered above
deal with two colorless operators, also separated by a small distance x.)
We will call such cases generically ``a small-size
color dipoles''. There would be
 three different kinds of those.

Historically the first example
are states of  heavy quarkonia.  The non-perturbative correction to
their energies  
was calculated by Voloshin and Leutwyler \cite{VL} by OPE:
$\delta E \sim  <0|g^2G_{\mu\nu}^2|0>
r^2 \ \tau$ where the spatial size $r\sim 1/(\alpha_s M)$ and the
rotational time $\tau \sim 1/(\alpha_s^2 M)$, both small for large
quark mass M. 
(I am not going to bother you with exact expression, in which
r appears as some dipole matrix element and $\tau$ as energy denominators,
as well by questions how small is  $\alpha_s$ and what is its argument.)

The question is whether this correction is indeed what is the case
in reality.
The answer probably exists or can be worked out using the spectroscopic
information  about Upsilons at hand: unfortunately
I am not aware of any
precision studies of it.

 Let me therefore move on to 
  another kind of
  ``dynamical dipoles'', the instantons, now 4-dimensionally symmetric,
with $r\sim \tau\sim\rho$.
One of the reason is I know it better, but there is also
a more scientific explanation of why one should do it instead.
Instantons in general
are much more sensitive tool, because  the tunneling probability
 contains the perturbative/semi-classic term (as well as  all non-perturbative corrections) 
{\it in the exponent}. 

(A side remark: as noticed in my paper \cite{Shu_95}, it is in particular should be
very important for
 fixing the  value of $\Lambda_{QCD}$. Experimentalists made tremendous
efforts to get it from scaling violation and similar effects containing
log($\Lambda_{QCD})$, but still get poor accuracy. Lattice practitioners
  use  hadronic masses and especially splittings of some 
 quarkonium levels for this purpose, which  are effects
 $const*\Lambda_{QCD}$, 
and improved the accuracy. But the
the density of small size instantons, 
calculated semi-classically in a seminal paper by 't Hooft \cite{tHooft},
contains  $O(\Lambda_{QCD})^b$
where $b=(11N_c/3-2N_f/3)\sim 9$.
 So
  the instanton-based determination of $\Lambda_{QCD}$ should
be potentially 10 times more accurate!)
 
 Similar to Voloshin-Leutwyler correction,
  the  OPE-based result \cite{SVZ} predicts the
 following correction to the density of instantons of size $\rho$:
\be 
dn(\rho)=dn_{pert} (\rho)(1+ {\pi^4\rho^4 \over 2 g^4} <0|g^2G_{\mu\nu}^2|0>+...) 
\label{e_cdg}
\ee
It is also second order dipole approximation: the instanton dipole is $O(\rho^2/g)$.
Note the generic  4-th power of  $\rho$
  is also determined by the dimension of the lowest gauge invariant local
operator.
 Note also the sign: it is nothing
else but a
generic attraction resulting from $any$ second order perturbation. 
However both these conclusions happen to be
in apparent conflict with  the lattice data (see below).

In   Fig.[1](a) we show  recent lattice data
for the  instanton size $distribution$ in pure
 SU(3) gauge theory
\cite{anna}.  (There are others but, this work  includes such refinements as
 improved
lattice action and back extrapolation to zero smoothening.)
 One can clearly see, that a rapid rise at small $\rho$
  turns into a strong suppression. 
The former behavior is consistent with the semi-classical one-loop
result \cite{tHooft}:
\be
\label{dist0}
{dN_0\over d\rho}|_{pert} = {C_{N_c} \over \rho^5}({8\pi^2 \over
  g^2(\rho)})^{2 N_c} (\rho \Lambda)^{b} 
\ee
where $C_{N_c}$ is the normalization constant, the $\rho^{-5}$ factor
and the term with the coupling constant comes
from the Jacobian
of the zero modes, and
$b=(11N_c/3-2N_f/3)$ as already mentioned.

\begin{figure}[ht]
\vskip .05in
 \begin{minipage}[c]{3.in}
 \centering
 \includegraphics[width=3.in]{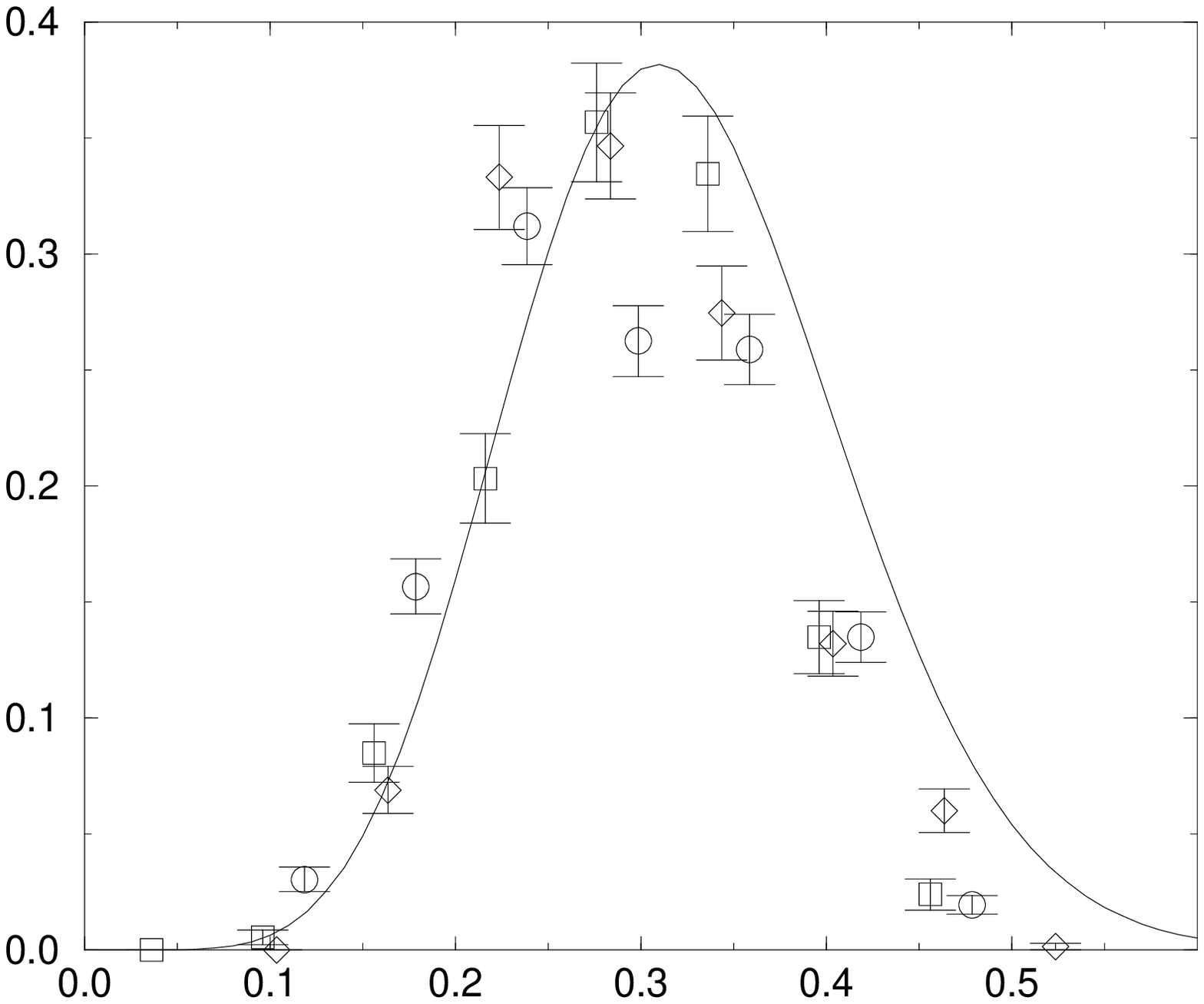}
 \end{minipage}
   \begin{minipage}[c]{3.in}
   \centering
\includegraphics[width=2.9in]{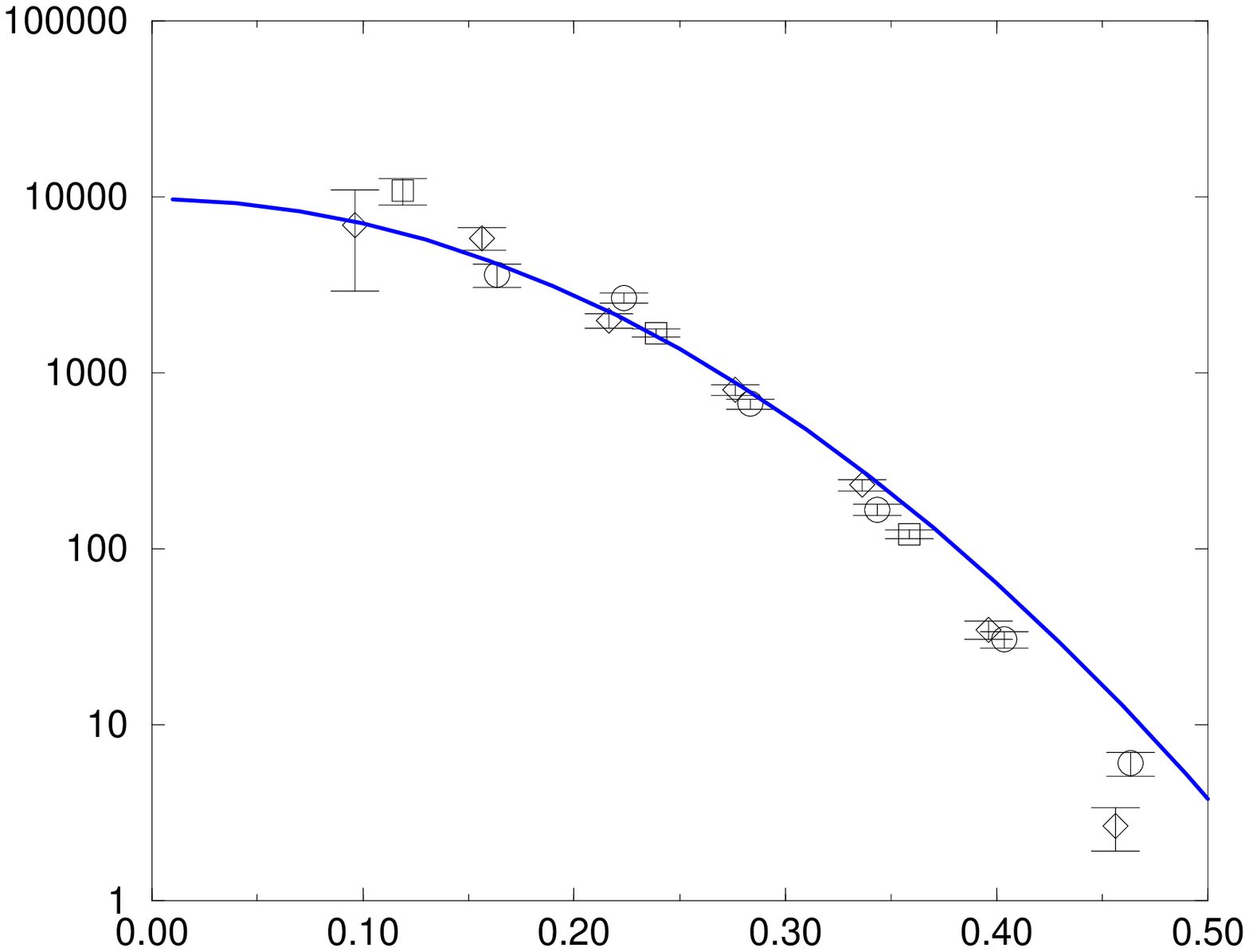}
 \end{minipage}
\vskip 0.1in 
  \vspace{-.05in}
  \caption{
(a) The instanton density $dn/d\rho d^4z$, [fm$^{-5}$] versus its size
 $\rho$ [fm]. (b) The combination  $\rho^{-6} dn/d\rho d^4z$, in which
 the main one-loop behavior drops out for $N_c=3,N_f=0$.
 The points are from the lattice work \protect\cite{anna},
for this theory, with 
$\beta$=5.85 (diamonds), 6.0 (squares) and 6.1 (circles). Their
comparison should demonstrate that results are
rather lattice-independent.
The line corresponds to the proposed
expression $\sim exp(-2\pi\sigma\rho^2)$, see text.
  }
\end{figure}

  Sharp maximum seen in   Fig.[1](a) appears at rather small
  $<\rho>\approx 1/3$ $fm$, much smaller that  their spacing
 $R\approx 1$ $fm$.  This results in a
  non-trivial  ``vacuum diluteness'' parameter  \cite{Shu_82} to which
  we turn in section 5,
 but
 now we are not interested in a typical instanton size
but rather in their suppression. Therefore
  we re-plot the same data  in  Fig.[1](b), with the leading
semi-classical behavior
taken away. Now the maximum is gone and
one finds the $same$ suppression
pattern  at both sides of the maximum.  
The OPE prediction (\ref{e_cdg}) is not seen: probably it is only true
at $very$ small $\rho$.
The suppression effect is clearly
$O(\rho^2)$, and not just for small $\rho$ but in the whole region.

\begin{figure}[t]
\vskip .05in
 \begin{minipage}[c]{3.in}
 \centering
 \includegraphics[width=2.9in]{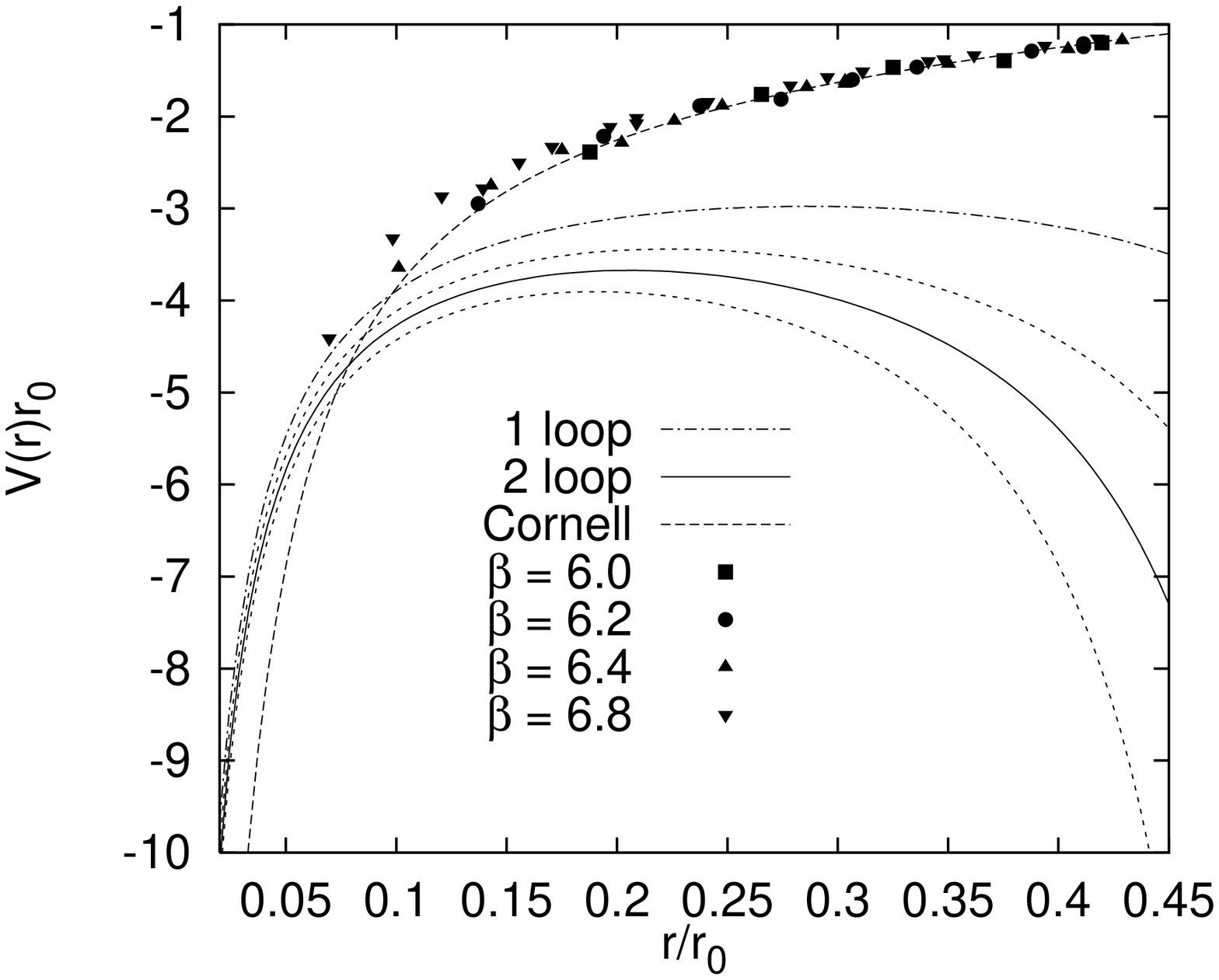}
 \end{minipage}
   \begin{minipage}[c]{3.in}
   \centering
\includegraphics[width=2.9in]{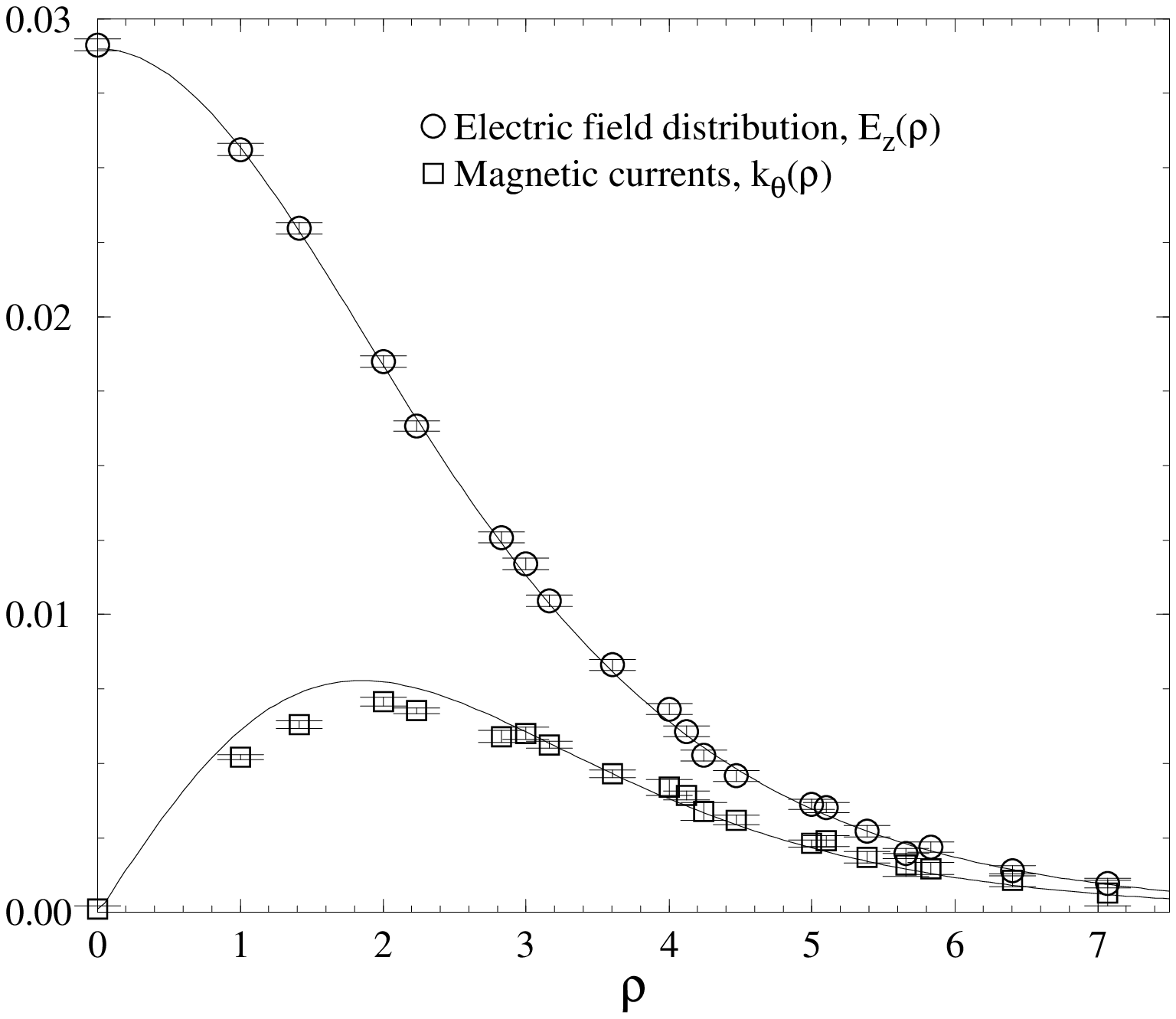}
 \end{minipage}
\vskip 0.1in
  \vspace{-.05in}
  \caption{
(a) The potential at small distances from \protect\cite{Bali_small},
the quantity $r_0\approx 0.5$~fm denotes the so called Sommer scale
  is used in lattice works. (b) The
 fits of the transverse distribution of electric and magnetic fields
in the string, obtained in lattice calculation, to ANO vortex solution
of the AHM \protect\cite{GIPS}. It leads to Higgs mass 1.3 GeV.
  }
\end{figure}

\subsection{Small Static Dipoles}
 Another way to look at these issues is to use    
{\em small static color dipoles}, or short strings. As  time 
scale $\tau$ is now unlimited, 
 there is no OPE prediction like (\ref{e_cdg}). However
  using just  the second order
 dipole approximation 
\cite{static_old} one gets a similar prediction of the non-perturbative
corrections to the static potential
$$
V(r) = -{4 \alpha_s(r) \over 3} { 1\over r} $$ \be
+ r^2 \int d\tau \ e^{(-{3 \alpha_s(r)\tau \over 2r})} <0|G_{\mu\nu}(\tau) U_\tau G_{\mu\nu}(0) U^+_\tau |0>)  
\ee
where the  field
strengths are separated by the time delay $\tau$, with
 $ U_\tau$  being the appropriate
parallel transports. Note that it is $O(r^2)$.
 
 However  recent  lattice data on V(r) 
 at small r \cite{Bali_small} have found
a clear O(r) effect instead (suggested previously in \cite{AZ}): 
\be
V(r) =- {4 \alpha_s(r) \over 3} { 1\over r} + \sigma_0 r +...  
\ee
The small-distance tension is larger than the asymptotic one
 $ \sigma_0\approx (4-5)  \sigma_\infty$. (I  however had to
warn the reader that it has rather uncertain error, because
 subtraction of the perturbative potential
depends on technical details.)

  This O(r) potential
for short strings makes sense to me mostly because there are many
other evidences that the QCD confining
strings  are  surprisingly thin.
  Lattice studies 
  \cite{Bali,GIPS} (the latter is shown in Fig.2b) show in particular that the
  so called ``energy radius'' (at which it decreases by 1/e) is about
$\delta_{1/e}\approx .18$ fm, while that for the
action distribution is about twice larger. 

This observation  also has many
phenomenological
consequences. One is just another argument explaining weak
string-string interactions known from Regge phenomenology. Another
 is ``hadron diluteness'': color  field inside hadrons
 occupy only few percent of the volume 
\be
(\delta_{1/e}/R_h)^2 \sim (1/5)^2
\label{dilute_hadron}
\ee
This is
contrary to the MIT bag model which views
the $whole$ hadronic interior to be in  the perturbative phase.
 In other words, the  value for the bag constant  $B_{MIT}\sim 50 \, MeV/fm^3$
was hugely under-estimated: it is $B\sim 1000  \, MeV/fm^3$ or more.
(Similar but different argument was
made two decades ago in \cite{Shu_78}.)

\section{Instanton-induced effects at small distances}
\subsection{The Renormalized Charge}

 We have not listed this issue in the list of observations presented
above, because the effective charge itself is not so easily observable.
Nevertheless, all calculations deal with it in some way or another,
and so it is impossible to avoid it in a theory discussion.
So let us ask what constitutes the first non-perturbative
correction to the effective charge.
It is defined here as the coefficient in an effective action
for some external weak background field $G_{\mu\nu}^a$,
in which $both$ perturbative loop corrections 
and the non-perturbative effects are to be included.

A very simple non-perturbative effect was suggested long ago by Callan,
Dashen and Gross (CDG) \cite{CDG}: instantons are simply dipoles and therefore
they are polarized by external field and produce some dielectric constant,
exactly like atoms do.  The external field is supposed to be normalized at some normalization scale
$\mu$, and CDG has proposed to include all instantons with size $\rho
<\rho_{max}=1/\mu$. The effective charge is then defined as:
\begin{equation} \label{eq_CDG}
{8\pi^2 \over g^2_{eff}(\mu)} = b \, \ln({ \mu\over \Lambda_{pert}}) 
\\ \nonumber
- {4\pi^2 \over (N_c^2-1)} \int^{\rho_{max}}_0 dn(\rho) \rho^4 
({8\pi^2 \over g^2_{eff}(\rho)})^2 
\end{equation}
  where 
$b=11N_c/3-2N_f/3$ is the usual one-loop coefficient of the beta function, and $dn(\rho)$ is the distribution of instantons 
(and anti-instantons) over  size.

\begin{figure}[ht]
\epsfxsize=4.in
\centerline{\epsffile{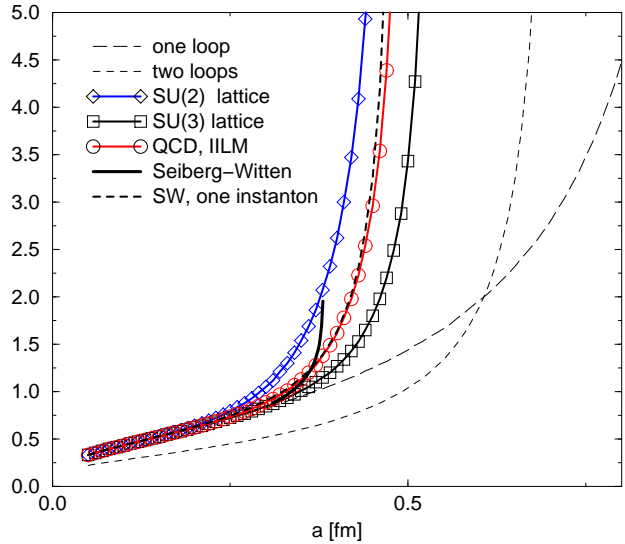}}
\vskip -0.05in
\caption[]{
 \label{fig.SW}
 The effective charge $b \,g^2_{eff}(\mu)/8\pi^2$ (b is the coefficient
of the one-loop beta function) versus normalization scale $\mu$ (in units of
its value at which the one-loop charge blows up). The thick solid line
correspond to exact solution \cite{SW} for the N=2 SYM, the thick dashed line
shows the one-instanton correction. Lines with symbols (as indicated on figure)
stand for N=0 QCD-like theories,
SU(2) and SU(3) pure gauge ones and QCD itself. Thin long-dashed and short-dashed lines are one and two-loop results.
}
\end{figure}

CDG could not calculate it then, as the instanton size distribution
was unknown. Plugging in lattice data (such as shown in Fig.1a)
for SU(2) gluodynamics, SU(3) gluodynamics we
get \cite{RRS}  curves shown in Fig.(\ref{fig.SW}). The curve for QCD with fermions
is not from lattice: we do not have it yet, so it corresponds to IILM.
Note a rapid departure from the perturbative logarithmic running
coupling constant. 

Is this indeed what is happening? In order to answer such questions
above
I have used either lattice data or hadronic phenomenology: let me now use
another tool from the theory tool-box, namely a {\em solvable model}.
So we    \cite{RRS} are going to compare
  the charge in  QCD to that  for the N=2
supersymmetric theory, for which the famous exact 
effective Lagrangian was found by Seiberg and Witten \cite{SW}.

\begin{equation}
{8\pi \over g^2(u)} = {K(\sqrt{1-k^2}) \over K(k)}
\end{equation}
where K is elliptic integral and the argument
\begin{equation}
k^2={(u-\sqrt(u^2-4 \Lambda^4))\over (u + \sqrt{u^2-4 \Lambda^4}))}
\end{equation}
is a function of gauge invariant vacuum expectation of squared scalar field 
\begin{equation}
u={1\over 2}<\phi^2>={a^2 \over 2} +{\Lambda^4 \over a^2} + \ldots
\end{equation}
and a is just its VEV. For large $a$ there is a weak coupling expansion
which includes  instanton effects  \footnote{It should
be noted that the first terms in this expansion have
been explicitly verified in instanton calculations \cite{inst}.}
\begin{equation} \label{pert}
{8\pi \over g^2(u)}={2 \over \pi} \left ( \log \left( {2 a^2 \over \Lambda^2 }\right) -
 {3 \Lambda^4 \over a^4}+ \ldots \right)
\end{equation}
The exact coupling blows up at $u=2 \Lambda^2$, which means that the factor
between the exact strong interaction scale and the perturbative
one is in this theory $\Lambda_{\infty}=2^{3/2}\Lambda$. Actually
this is the ratio of the scale $\sqrt{u}$ to the scale $a$
at which the perturbatively evolved coupling (one-loop) blows up.
If one were to account for the next term in the expansion of $u$,
the ratio of scales is reduced to $\sqrt{2} \sqrt{2+\sqrt{2}}$.
The fact that instanton effects can be important at such a high
scale was anticipated in Ref. \cite{CDG} and is presumably due
to the significance of the pre-factor in instanton calculations.)

The behavior is shown in  Fig.(\ref{fig.SW}), where we
have included both a curve which shows the full 
coupling (thick solid line), as well as a curve which illustrates
only the one-instanton correction (thick dashed one).
The units on both axis are chosen so that the naive one-loop
running looks the same in both theories. And -- surprise, surprise --
the instanton-induced corrections seem to be the nearly the same as
well.

I think this comparison teaches us few lessons. The main of them: if
the instanton-induced corrections to the charge itself
becomes large, the perturbative expansion in g have to  be
abandoned. I do not know what exactly one can make out of (nearly  ideal)
numerical matching
of these two theories, QCD and N=2 SUSY gluodynamics.
Note however the very rapid change of the coupling induced by
instantons,
although different in formulae, but so similar in the curves.
It is also of interest that the full multi-instanton sum makes
the rise in the coupling even more radical than with only the
one-instanton correction incorporated. It is also interesting
to observe that at the scale where the true coupling blows up,
the perturbatively evolved coupling is still not very large.
 Individually, the perturbative log and instanton corrections
are well defined at this region: however they cancel each other in the
inverse charge. This
is encouraging from the point of view of developing a consistent
expansion for the
instanton corrections. The rapid rise in the coupling is also
encouraging in that it ensures that perturbation theory is valid
almost to the point where it blows up. For a consistent
picture of QCD, in which perturbation theory still appears to
be applicable at the $c$-quark scale, while the theory
is non-perturbative at 1 
GeV, such a dramatic effect is essential.

\subsection{The Chiral Scale in Spin-0 Light-Quark correlators}
  The physical origin of the chiral scale has been
traced down to {\em instantons-induced effects} \cite{Shu_82}, for recent detailed
review  see
\cite{SS_98}. The main points are: (i) 
 the instanton-induced interaction strong and short-range, (ii) it
is attractive for the
pion channel, repulsive for the $\eta'$, and (iii) (to first order) it does 
not affect the vector channels such as $\rho,\omega$. It 
explains $both$  successes and failures
of the OPE sum rules, reproducing the
   phenomenology also for all spin-0  correlation functions
  \cite{Shu_83}.
So far, we have only
  mentioned the pion contribution, and noted that it deviates from
perturbative behavior at unusually small distances.

  After this prelude we return  to evaluation of
 the short distance behavior 
of correlation functions in the single-instanton approximation (SIA) 
\cite{Shu_83}. The main idea is that if $x-y$ is small compared to 
the typical instanton separation $R$, we expect that the contribution 
from the instanton $I=I_*$ closest to the points $x$ and $y$ will dominate
over all others. For quark propagator in the subspace of the instanton
zero modes this 
implies 
\be
\label{S_SIA}
S(x,y)\; = \;\sum_{IJ}\psi_I(x)\left(\frac{1}{T+im}\right)_{IJ}\!\!
        \psi^\dagger_J(y)
      \;\simeq\;\psi_{I_*}(x)\left(\frac{1}{T+im}\right)_{I_*I_*}
      \!\!\psi^\dagger_{I_*}(y)
      \;\simeq\;\frac{\psi_{I_*}(x)\psi^\dagger_{I_*}(y)}{m^*}
\ee
where we have approximated the diagonal matrix element by its average, 
$(T+im)^{-1}_{I_*I_*}\simeq N^{-1}\sum_I (T+im)^{-1}_{II}$, and introduced 
the effective mass $m^*$  $(m^*)^{-1}
=N^{-1}\sum \lambda^{-1}$. In the following we will use the mean field 
estimate $m^*=\pi\rho(2n/3)^{1/2}$. As a result, the propagator in the 
SIA looks like the zero mode propagator of a single instanton, but for 
a particle with an effective mass $m^*$. 

  The $\pi$ and $\eta'$ correlators receive zero modes contributions.
In the single instanton approximation, we find \cite{Shu_83}
\be
\label{ps_SIA}
\Pi^{SIA}_{\pi,\eta'}(x) &=& \pm\int d\rho\, n(\rho)
 \frac{6\rho^4}{\pi^2}\frac{1}{(m^*)^2}
 \frac{\partial^2}{\partial (x^2)^2}
 \left\{ \frac{4\xi^2}{x^4} \left(
 \frac{\xi^2}{1-\xi^2} +\frac{\xi}{2}
 \log\frac{1+\xi}{1-\xi}\right)\right\},
\ee
where $\xi^2=x^2/(x^2+4\rho^2)$. There is also a non-zero mode contribution 
to these correlation functions which is not very important.

 Putting in standard instanton
parameters, we find that  at very short distances
$x\simeq .3$ fm, this contributions to the correlator coming 
from the instanton zero modes are larger than the  contributions 
included in the OPE (see Fig.4a). 

For the $\eta'$, the instanton contribution is strongly repulsive.
The flip of the sign is easy to understand:
the instanton corrections comes only from flavor-changing
diagram $\bar u u \rightarrow \bar d d$, which has different sign in
correlators
of $J_{\pi0}=(\bar u \gamma_5 u-\bar d \gamma_5 d)$ and  $J_{\eta'}=(\bar u \gamma_5 u+\bar d \gamma_5 d)$
This is because  the  instanton  accounts for the $U(1)_A$
anomaly. The SIA was extended to the full pseudoscalar nonet in 
\cite{Shu_83}. It was shown that simply replacing $m^*\to m^*+m_s$
gives a good description of $SU(3)$ flavor symmetry breaking and 
the $\pi,K,\eta$ correlation functions. 

  In the vector $\rho,\omega$  correlators the zero modes cannot contribute since 
the chiralities do not match. Non-vanishing contributions come from 
the non-zero mode propagator  and from interference between 
the zero mode part and the leading mass correction 
\be
\Pi^{SIA}_\rho(x,y) &=& {\rm Tr}\left[ \gamma_\mu S^{nz}(x,y)
\gamma_\mu S^{nz}(y,x) \right] + 2{\rm Tr}\left[ \gamma_\mu
\psi_0(x)\psi_0^\dagger(y)\gamma_\mu \Delta(y,x)\right]
\ee
The latter term survives even in the chiral limit, because the factor 
$m$ in the mass correction is cancelled by the $1/m$ from the zero mode.
Also note that the result corresponds to the standard DIGA, so 
multi-instanton effects are not included. After averaging over the 
instanton coordinates, we find \footnote{There is a mistake by an
overall factor 3/2 in the original work.} \cite{AG_78}
\be
\label{vec_SIA}
\Pi^{SIA}_\rho(x) &=&  \Pi^0_\rho + \int d\rho\, n(\rho)
 \frac{12}{\pi^2}\frac{\rho^4}{x^2}\frac{\partial}{\partial (x^2)}
 \left\{ \frac{\xi}{x^2} \log\frac{1+\xi}{1-\xi}\right\}
\ee
Similar to the OPE, 
this correction
  is only weakly attractive at intermediate distances.

   An interesting observation is the fact that it
is {\em the only} singular term in the instanton contribution. In fact,
the OPE of {\em any} mesonic correlator in {\em any} self dual field 
contains only dimension 4 operators \cite{DS_81}. This means that for 
all higher order operators either the Wilson coefficient vanishes 
(as it does, for example, for the triple gluon condensate $\langle 
g^3f^{abc}G^a_{\mu\nu} G^b_{\nu\rho}G^c_{\rho\mu}\rangle$) or the 
matrix elements of various operators of the same dimension cancel 
each other. This is a very remarkable result, because it 
helps to explain the success of QCD sum rules based on the OPE
in a number of channels. In the instanton model, the gluon fields 
are very inhomogeneous, so one would expect that the OPE fails 
for $x\ll \rho$. The Dubovikov-Smilga result shows how many observables
can be absolutely blind to very strong gauge fields, as long as 
they are locally self dual.

\subsection{Instantons and the  duality scale for $J^{PC}=O^{\pm}$ glueballs}
   Let me first explain qualitatively why in this case the scale
   should be
   larger than in the light quark channels. 
The reason is the gluonic field strength inside instantons is very large.

The magnitude of the
color fields inside the instanton is $O(1/g)$, and  in the
correlation functions in question (such as $<G^2(x) G^2(y)>$) it
enters in the 4-th power. Therefore, the instanton correction 
is proportional to the action of the instanton squared,
$(8\pi^2/g^2)^2\sim 100$ where for estimate we have used the action to
be of the order of 10. The corresponding correction with light quarks
have zero modes instead of gluonic field strength, and its integral
over space is not the action but just one, from the normalization
of fermionic mode. So we speak about enhancement of roughly 2 orders
of magnitude.

   For one-instanton calculation we can start from
expansion of  the gluon operators
around the classical fields.
In the lowest order one should simply 
 insert the classical field of an instanton into the operators, it
 leads to 
 \cite{Shu_83}
\be
\label{gb_SIA}
\Pi_{S,P}^{SIA}(x)&=&  \int d\rho\, n(\rho)
 \frac{8192\pi^2}{g^4\rho^4}\frac{\partial^3}{\partial (x^2)^3}
 \left\{ \frac{\xi^6}{x^6} \left( \frac{10-6\xi^2}{(1-\xi^2)^2}
 +\frac{3}{\xi}\log\frac{1+\xi}{1-\xi} \right)\right\}
\ee
where $\xi$ is defined as in (\ref{ps_SIA}). There is no classical 
contribution in the tensor channel, since the stress tensor in the 
self-dual field of an instanton is zero. Note that the perturbative 
contribution in the scalar and pseudoscalar channels have opposite sign, 
while the classical contribution has the same sign. To first order in the 
instanton density, we therefore find all  three scenarios :
 {\em attraction} in the scalar channel, 
{\em repulsion} in the pseudoscalar and {\em no} effect in the tensor 
channel. 

The single-instanton prediction can be compared with the OPE,
and it is much larger indeed. Furthermore, if one uses
the standard $\rho, n$ of the instanton vacuum, it matches it
the scale inferred from low energy theorem very well.

   Numerical calculations of glueball correlators in different 
instanton ensembles were performed in \cite{SS_95}. At short distances, 
the results are consistent with the single instanton approximation. At 
larger distances, the scalar correlator is modified due to the presence
of the gluon condensate. This means that (like the $\sigma$ meson), the
correlator has to be subtracted and the determination of the mass is 
difficult. In the pure gauge theory we find $m_{0^{++}}=1.5$ GeV.
While the mass is consistent
with QCD sum rule predictions, the coupling is much larger than expected
from calculations that do not enforce the low energy theorem.

   In \cite{SS_95} we also measured glueball wave functions. The most
important result is that the scalar glueball is indeed small, $r_{0^{++}}
= .2$ fm 
and determined by the size of an instanton,
 while the tensor is much bigger,
sensitive to interactions at the size  determined
by the average distance between instantons 1 fm.

   In the pseudoscalar channel the correlator is very repulsive and there
is no clear indication of a glueball state. In the full theory (with 
quarks) the correlator is modified due to topological charge screening.
The non-perturbative correction changes sign and a light (on the scale
of the glueballs) state, the $\eta'$ appears. Non-perturbative corrections
in the tensor channel are tiny. Isolated instantons and anti-instantons 
have a vanishing energy momentum tensor, so the result is entirely due 
to interactions. 

\section{Confinement effects at small distances}
 \subsection{Instanton Suppression}
The central idea of this Section is  
 that  $O(\rho^2)$ suppression of instantons can be due to
a   ``dual superconductivity'' \cite{dual},  a scenario in which  some
composite objects 
condense, forming the non-zero
vacuum expectation value (VEV) of the magnetically charged
 scalar field $\phi$. 
 My first (naive)
argument was that in such theory, unlike the QCD itself,
  at least there is the dimension-2 operator $|\phi|^2$. 

The composites may  be
magnetic  monopoles \cite{dual,SW}, or  P-vortices, or something else:
anyway 
one is lead to an
 incarnation of the old 
Landau-Ginzburg effective theory,  Abelian Higgs Model (AHM),
describing interaction of a
``dual photon'' and ``dual Higgs''  fields. 
AHM was
applied  to the description of 
 the QCD strings, as    
Abrikosov-Nielsson-Olesen  vortices \cite{ANO}.

Before we go into details,
let us point out a striking similarity
between these two problems. A vortex  is the 2d topologically
non-trivial 
configuration, in which $\phi$ vanishes at the
center, the Dirac string where the dual potential must be singular. 
An instanton problem is in a way the previous one squared. The 
4d  picture of the fields is like two string cross sections
in two orthogonal 2d planes. Higgs field
$\phi$ again vanishes at the
center,  because  in the singular
gauge
(the only one good for multi-instanton configurations)
 the gauge field is $A_\mu(x)^2 \sim 1/x^2$  at the origin,
acting as a centrifugal barrier.
Since ``melting'' of the dual superconductor at the center
 is not a small
modification, one generally cannot expect the OPE-type calculations to
hold.
In
both problems one has first to solve for the field and then calculate
the energy or action. Fortunately, for instantons in a  Higgsed vacuum
it was already done by 't Hooft \cite{tHooft}: for fundamentally
charged
Higgs the answer is
\be
\Delta S= 4\pi^2\rho^2 |<\Phi>|^2
\label{tHooft_corr}
\ee
 Note that it leads to the $O(\rho^2)$ suppression law
we need to explain  Fig.[1](b), and that $\Delta S$
 should not necessarily be small.
We return to speculations on the exact nature of the Higgs and the
 dual photon fields of the Landau-Ginzburg model (needed to evaluate
 the
strength of the effect) below.

Encouraged by this, we return to instantons and
 try to apply the same reasoning.
Since both  the 't Hooft correction (\ref{tHooft_corr}) and the string tension
(\ref{tension}) scales as the Higgs VEV squared, we expect qualitatively
that 
\be
\label{suppr}
{dN\over d\rho} = {dN\over d\rho}|_{pert} exp{(-C\sigma\rho^2)}
\ee
where C is some numerical constant.
 In order to find C one has to identify the scalar and the dual photon fields
of the AHM, and explain how they
are coupled to  the colored gauge field of the instanton.
In the AHM treatment of the QCD string \cite{Baker_etal}
the magnetic field of the dual photon is 
identified directly with the color-electric gauge field
inside the string. It does not create problems because this electric field can
be considered Abelian.
 
 Applying the same ideas for instantons, let us first note that their 
 $self-duality$ helps: because  electric and
magnetic fields are identical, the ``magnetic'' potential $C_\mu$ and the
original one $A_\mu$ are also the same. However both are intrinsically
non-Abelian, so only a particular component (or a combination of
those)
 can be identified with the Abelian ``dual photon''  of the effective theory.
In other words, an {\em Abelian
projection} is inevitable, and there is no unique or preferred way to do it. 
Lacking better ideas, we
 simply do what lattice people do: just select one of possible projections
and see what happens. Clearly, selecting Higgs field interacting with
a particular component of the gauge potential means breaking the gauge
group 
(which the vacuum of the Standard model does and that of QCD does not).
 But we proceed anyway, simply  re-scaling the dual fields in a way 
that their Lagrangian (1) matches the 't Hooft one. If we do so,
 it leads to identification
$<\Phi>^2=(2/3)\phi_0^2$, or the constant C in (\ref{suppr})
to be $C=2\pi$. \setcounter{footnote}{0}
Putting it all together, we can now compare\footnote{ The standard string tension value
$\sigma=(450 MeV)^2$   is  traditionally used to set absolute scale
of the lattice data for non-physical theories like for 
the pure SU(3) gauge theory simulations we
used: so the instanton sizes expressed in fm is consistent with it.}
this result to the exponential suppression. The corresponding 
 curves are shown 
 in Fig.1, and it works very well. 

{\bf Brief summary}. The ``dual superconductivity'' leading to
confinement in the QCD vacuum seems to be surprisingly
robust. Instanton suppression  and/or string tension
considered before both fix the AHS {\em Higgs VEV} rather accurately.
The string size and small-r potential provide hints that
the {\em Higgs} and  {\em dual photon masses} are large, in the 1-1.5 GeV range.
\setcounter{footnote}{0}
Their exact nature remains unclear\footnote{
 So the status of AHM Higgs  is, ironically,  not that different from
Higgs particle of the Standard Model.}.

{\bf Outlook}: one may  test our suggestions by comparing instanton suppression
in  theories
with
variable number of colors and/or flavors. 
Unfortunately available data
for the SU(2) color group or SU(3) with dynamical
fermions are not yet good enough to do so.

Another challenging set of questions is related with
instanton  suppression 
 at non-zero temperatures T and/or densities. 
 At high T or density, in the quark-gluon
plasma phase, the answer is clear: the instanton
 electric fields are again  suppressed, but now by the usual
Debye screening\cite{Shu_78}. It leads to a factor $ exp(-a(T,\mu)
\rho^2)$ 
similar to the one discussed above, where the coefficient
$a(T,\mu=0)$ was calculated in  \cite{PY} and then generalized to the
$\mu\neq 0$ in
\cite{Shu_82}. Note that confinement is not completely gone: it
remains for spatial Wilson loops.
 The most interesting point is what happens
close to the deconfinement transition. Since 2 and 3 color gouge
theories
have it of the second and the first kind, respectively, a detailed
study of the instanton suppression at $T\approx T_c$ is of great interest.

\subsection{Long and Short Strings}
 Now we briefly review
  applications of the dual superconductivity idea to the QCD
 string, or  the 
ANO vortex line, done in
 a series of papers  \cite{dualmodel}.
 Among clear successes of this approach is: (i) 
prediction of  weak string-string interaction, putting it around the
boundary of type I and II superconductivity;  (ii) prediction of
a whole set of potentials other than central.
 Both agree well with available lattice data, 
for a review see \cite{Bali}.

The effective Lagrangian used in  \cite{Baker_etal} is
\be L={4\over 3} [{1\over 4} (\partial_\mu C_\nu - \partial_\nu C_\mu)^2 + \ee
$$ {1\over 2} | (\partial_\mu - ig_m C_\mu)\phi |^2 +
{\lambda \over 4} ( |\phi|^2-|\phi_0|^2)^2   ]
$$
where we have omitted interaction with quarks at the ends.
$C_\mu$ is dual color potential coupled to Higgs with magnetic
coupling $g_m=2\pi/g$.
Assuming that we are exactly at
the boundary of the type I and II superconductivity,
the masses of the Higgs and the ``dual photon'' are
 equal $M_\phi=M_C=g_m \phi_0  $. The 
(classical) string tension is directly related to Higgs VEV
\be
\sigma={4\pi\over 3} |\phi_0 |^2
\label{tension}
\ee

 In effective dual model \cite{Baker_etal}
 the  string width  is related to masses of dual photon and Higgs,
being the {\em large non-perturbative scale} of the 3-ed kind we are speaking about.
 The data mentioned put it in the
``glueball mass  range'',  around 1.3 GeV according to  Fig.2b
(It is difficult for me to access the error involved.)

\section{QCD Vacuum as the Instanton Liquid}
\subsection{The qualitative picture}
 It was already mentioned that, as pointed out  in \cite{Shu_82},
  the typical instanton size is significantly smaller than their
 separation $R=n^{-1/4}\approx 1 fm$, or
\be 
\label{eq_rho}
\rho_{\rm max}&\sim& 1/3\,{\rm fm} .
\ee
If so,  the following qualitative
picture of the QCD vacuum emerges
\begin{enumerate}

\item Since the instanton size is significantly smaller than the typical 
separation $R$ between instantons, $\rho/R \sim 1/3$, the vacuum is fairly 
dilute. The fraction of spacetime occupied by strong fields is only a few 
per cent.

\item The fields inside the instanton are very strong $G_{\mu\nu}\gg\Lambda_{QCD}^2$.
This means that the semi-classical approximation is valid, and the typical 
action is large
\be
\label{S_typ}
 S_0 = 8\pi^2/g^2(\rho) \sim 10-15 \gg 1 .
\ee
Higher order corrections are proportional to $1/S_0$ and presumably small.

\item Instantons retain their individuality and are not destroyed by 
interactions. From the dipole formula, one can estimate
\be
\label{S_int_typ}
 |\delta S_{int}| \sim (2-3) \ll S_0  .
\ee

\item Nevertheless, interactions are important for the structure of the 
instanton ensemble, since
\be
  \exp|\delta S_{int}| \sim 20 \gg 1 .
\ee
This implies that interactions have a significant effect on correlations among 
instantons, the instanton ensemble in QCD is not a dilute gas, but an 
interacting liquid. 
\end{enumerate}

 The aspects of the QCD vacuum for which the instantons are most
 important are
those related with
light fermions.
Their importance  in the context of chiral symmetry breaking is 
related to the fact that the Dirac operator has a chiral zero mode in the 
field of an instantons. 
These zero modes are  localized quark states around instantons, like
atomic states of electrons around nuclei. At finite density of the instantons
those states can become collective, like atomic states in metals. 
The resulting de-localized state corresponds to the wave function of the quark
condensate. 

\subsection{ How to calculate?}
\label{sec_how}
In this lectures there is no place for technical details, which may be found 
in the original papers. Rather we try to explain the logical structure and 
possible alternatives of the theory. For that reason, discussion of possible 
methods to calculate something with instantons is put into the introduction, 
with only results of the calculation reported below.

In the QCD partition function there are two types of fields, gluons and quarks,
and so the first question one addresses is which integral to take first.

(i) One way is to eliminate $gluonic$ degrees of freedom first. Physical 
motivation for that may be that gluonic states are heavy and an effective 
fermionic theory should be better suited to derive an effective low-energy
theory. In the instanton framework, this approach was started by 't Hooft who 
discovered that instantons lead to new effective interactions between light 
quarks. We will present the explicit form of such four-fermion interaction for 
two-flavor QCD below. 

It is a well-trotted path and one can follow it along the development for 
a similar four-fermion theory, the NJL model. One can do simple mean field or 
random field approximation (RPA) diagrams, and find the mean condensate and 
properties of the mesons. Unfortunately, it is difficult to do more. For 
example, baryons are states with three quarks, and using quasi-local 
four-fermion Lagrangians for the three body problem is technically a very 
difficult (although solvable) quantum mechanical problem. There were no 
attempts to sum more complicated diagrams.

(ii) The opposite strategy can be to do $fermion$
 integral first. It is a simple 
step, because they only enter quadratically, leading to a fermionic determinant.
This is the way lattice people proceed. In the instanton approximation, it leads
to the Interacting Instanton Liquid Model, defined by the following partition 
function:
\be 
\label{part_fct} 
Z =   \sum_{N_+,\, N_-} {1 \over N_+ ! N_- !}\int 
    \prod_i^{N_+ + N_-} [d\Omega_i\; d(\rho_i) ] 
    \exp(-S_{\rm int})\prod_f^{N_f} \det(\hat D+m_f) \, , 
\ee 
describing a system of pseudo-particles interacting via the bosonic action and 
the fermionic determinant. Here $d\Omega_i=dU_i\, d^4z_i\, d\rho_i$ is the 
measure in color orientation, position and size, associated with single 
instantons and  $d(\rho)$ is the single instanton density 
$d(\rho)= dn_{I,\bar I}/d\rho dz$.
 
The gauge interaction between instantons is approximated by a sum of pure 
two-body interaction $S_{\rm int}=\frac{1}{2}\sum_{I\neq J}S_{\rm int} 
(\Omega_{IJ})$. Genuine three body effects in the instanton interaction are not
important as long as the ensemble is reasonably dilute. Implementation of this 
part of the interaction (quenched simulation) is quite analogous to usual 
statistical ensembles made of atoms. 
 
As already mentioned, quark exchanges between instantons are included in the 
fermionic determinant. Finding a diagonal set of fermionic eigenstat \-es of the 
Dirac operator is similar to what people are doing, e.g., in quantum chemistry
when electron states for molecules are calculated. The difficulty of our 
problem is however much higher, because this set of fermionic states should be
determined for $all$ configurations which appear during the Monte-Carlo process.

If the set of fermionic states is however limited to the subspace of instanton 
zero modes, the problem becomes tractable numerically. Typical calculations in 
the IILM involved up to N$\sim 100$ instantons (+anti-instantons):
 it means that 
the determinants of $N\times N$ matrices are involved. Such determinants can be 
evaluated by the ordinary workstation (and even PC these days) so quickly, that 
straightforward Monte Carlo simulation of IILM is possible in a matter of 
minutes. On the other hand, expanding the determinant in a sum of products of 
matrix elements, one can easily identify the sum of all closed loop diagrams up 
to order $N$ in the 't Hooft interaction. Thus, in this way we take care of 
(practically) all orders!
 
\subsection{Correlation functions}
\begin{figure}[b!]
\vspace{-3mm}
\centerline{\epsfxsize=10cm\epsffile{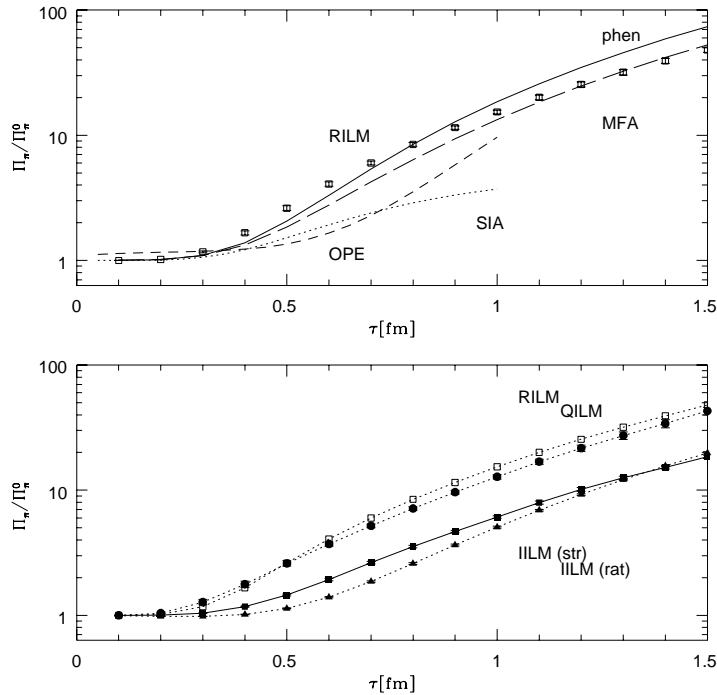}}
\vspace{-3mm}
\caption{\label{fig_pi_cor}
Pion correlation function in various approximations and instanton ensembles. In
the top figure we show the phenomenological expectation (solid), the OPE 
(dashed), the single instanton (dash-dotted) and mean field approximations 
(dashed) as well as data in the random instanton ensemble. In the bottom figure
we compare different instanton ensembles, random (open squares), quenched 
(circles) and interacting (streamline: solid squares, ratio ansatz solid 
triangles).}
\end{figure}
  
  There is no place here to explain the details of how we performed calculations in
  the
Interacting Instanton Liquid Model (IILM): they are done by Monte
Carlo, in a way not so different from standard methods used in other
statmech
models. Different versions of the model (mentioned in figures below as
IILM(rat) etc) differ by a particular ansatz for gauge field used,
from which the interaction is calculated. Note also, that these figures
mostly contain also a curve marked ``phen'': this is how the
correlator actually look like, according to phenomenology.

\begin{figure}[t!]
\centerline{\epsfxsize=10cm\epsffile{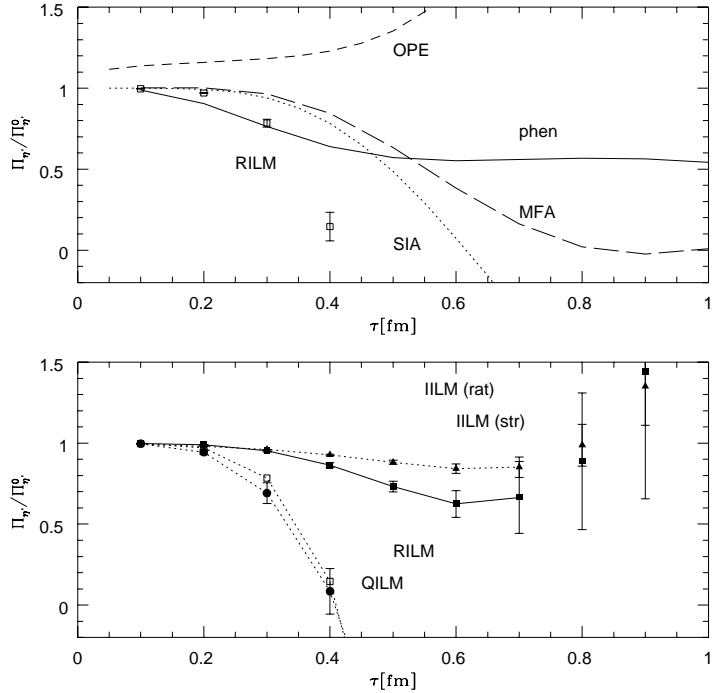}}
\vspace{-3mm}
\caption{\label{fig_eta_cor}
Eta prime meson correlation functions. The various curves and data sets are
labeled as in Fig.~\ref{fig_pi_cor}.}
\end{figure}

We simply jump to show a sample of results. 
Correlation functions in the different instanton ensembles were
calculated
in several papers (see original refs in \cite{SS_98}).
Some of them (like vector and axial-vector ones) turned out to be
easy:
nearly any variant of the
instanton model can reproduce well the (experimentally
known!)
correlators. Some of them are sensitive to details of the model very much:
two such cases are shown in Figs.~\ref{fig_pi_cor}-\ref{fig_eta_cor}. The pion 
correlation functions in the different ensembles are qualitatively very similar.
The differences are mostly due to different values of the quark condensate 
(and the physical quark mass) in the different ensembles. Using the Gell-Mann, 
Oaks, Renner relation, one can extrapolate the pion mass to the physical value 
of the quark masses. The results are consistent with the experimental value in 
the streamline ensemble (both quenched and unquenched), but clearly too small 
in the ratio ansatz ensemble. This is a reflection of the fact that the ratio 
ansatz ensemble is not sufficiently dilute.

\begin{figure}[b!]
\vspace{-3mm}
\centerline{\epsfxsize=10cm\epsffile{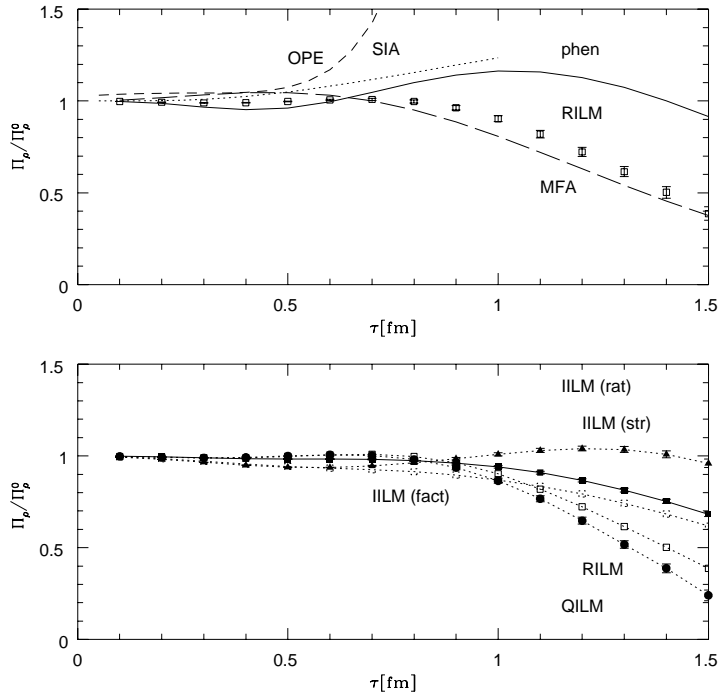}}
\vspace{-3mm}
\caption{\label{fig_rho_cor}
Rho meson correlation functions. The various curves and data sets are labeled
as in Fig.~\ref{fig_pi_cor}. The dashed squares show the non-interacting part
of the rho meson correlator in the interacting ensemble.}
\end{figure}

In Fig.~\ref{fig_rho_cor} we also show the results in the $\rho$ channel. The 
$\rho$ meson correlator is not affected by instanton zero modes to first order 
in the instanton density. The results in the different ensembles are fairly 
similar to each other and all fall somewhat short of the phenomenological 
result at intermediate distances $x\simeq 1$ fm. We have determine the $\rho$ 
meson mass and coupling constant from a fit. The $\rho$ meson mass is somewhat 
too heavy in the random and quenched ensembles, but in very good agreement 
with the experimental value $m_\rho=770$ MeV in the interacting ensemble.

Since there are no interactions in the $\rho$ meson channel in the RPA, it is 
important to study whether the instanton model provides any binding at all. In 
the instanton model, there is no confinement, and $m_\rho$ is close to the two 
(constituent) quark threshold. In QCD, the $\rho$ meson is also not a true bound
state, but a resonance in the 2$\pi$ continuum. In order to determine whether
the continuum contribution in the instanton model is predominantly $2\pi$ or 
2 quark would require the determination of the corresponding three point 
functions (which has not been done yet). Instead, we will compare the full 
correlation function with the non-interacting one, where we use the average 
(constituent quark) propagator determined in the same ensemble 
(Fig.~\ref{fig_rho_cor}). As explained above, this comparison provides a measure
of the strength of interaction. We observe that there is an attractive 
interaction generated in the interacting liquid. The interaction is due to 
correlated instanton-anti-instanton pairs. This is consistent with the fact 
that the interaction is considerably smaller in the random ensemble. In the 
random model, the strength of the interaction grows as the ensemble becomes 
more dense. However, the interaction in the full ensemble is significantly 
larger than in the random model at the same diluteness. Therefore, most of the 
interaction comes from dynamically generated pairs.

The situation is drastically different in the $\eta'$ channel. Among the $\sim 
40$ correlation functions calculated in the random ensemble, only the $\eta'$ 
and the isovector-scalar $\delta$ were found to be completely
unacceptable: The correlation function decreases very rapidly and becomes 
$negative$ at $x\sim 0.4$ fm. This behavior is incompatible even with a normal 
spectral representation. The interaction in the random ensemble is too 
repulsive, and the model ``over-explains" the $U(1)_A$ anomaly. 

The results in the unquenched ensembles (closed and open points) significantly 
improve the situation. This is related to dynamical correlations between 
instantons and anti-instantons (topological charge screening). The single 
instanton contribution is repulsive, but the contribution from pairs is 
attractive. Only if correlations among instantons and 
anti-instantons are sufficiently strong, the correlators are prevented from 
becoming negative. Quantitatively, the $\delta$ and $\eta_{\rm ns}$ masses in 
the streamline ensemble are still too heavy as compared to their experimental 
values. In the ratio ansatz, on the other hand, the correlation functions even 
show an enhancement at distances on the order of 1 fm, and the fitted masses 
are too light. This shows that the $\eta'$ channel is very sensitive to the 
strength of correlations among instantons.

In summary, pion properties are mostly sensitive to global properties of the 
instanton ensemble, in particular its diluteness. Good phenomenology demands 
$\bar\rho^4 n\simeq 0.03$, as originally suggested in\cite{Shu_82}. The 
properties of the $\rho$ meson are essentially independent of the diluteness, 
but show  sensitivity to $\bar I I$ correlations. These correlations become 
crucial in the $\eta'$ channel.

\subsection{Baryonic correlation functions}
\label{sec_bar_cor}

 Existence of a 
strongly attractive interaction in the pseudoscalar quark-antiquark (pion) 
channel also implies an attractive interaction in the scalar quark-quark 
(diquark) channel. This interaction is phenomenologically very desirable, 
because it immediately explains why the nucleon and lambda are light, while 
the delta and sigma are heavy.

The proton currents (with no derivatives
and the minimum number of quark fields) with positive parity and spin $1/2$ are
given by\cite{Iof_81}. 
It is  useful to rewrite these currents in terms of scalar and 
pseudoscalar diquarks
\be
\label{ioffe_ps}
\eta_{1,2} &=& (2,4) \left\{ \epsilon_{abc} (u^a C d^b)\gamma_5 u^c 
 \mp \epsilon_{abc} (u^a C\gamma_5 d^b) u^c \right\}.
\ee
Nucleon correlation functions are defined by $\Pi^N_{\alpha\beta}(x)=\langle 
\eta_\alpha(0)\bar\eta_\beta(x) \rangle$, where $\alpha,\beta$ are the Dirac 
indices of the nucleon currents. In total, there are six different nucleon 
correlators: the diagonal $\eta_1\bar\eta_1,\,\eta_2\bar\eta_2$ and 
off-diagonal $\eta_1\bar\eta_2$ correlators, each contracted with either the 
identity or $\gamma\cdot x$. Let us focus on the first two of these correlation
functions (for more detail, see\cite{SS_98} and references therein).

\begin{figure}[htb]
\centerline{\epsfxsize=10cm\epsffile{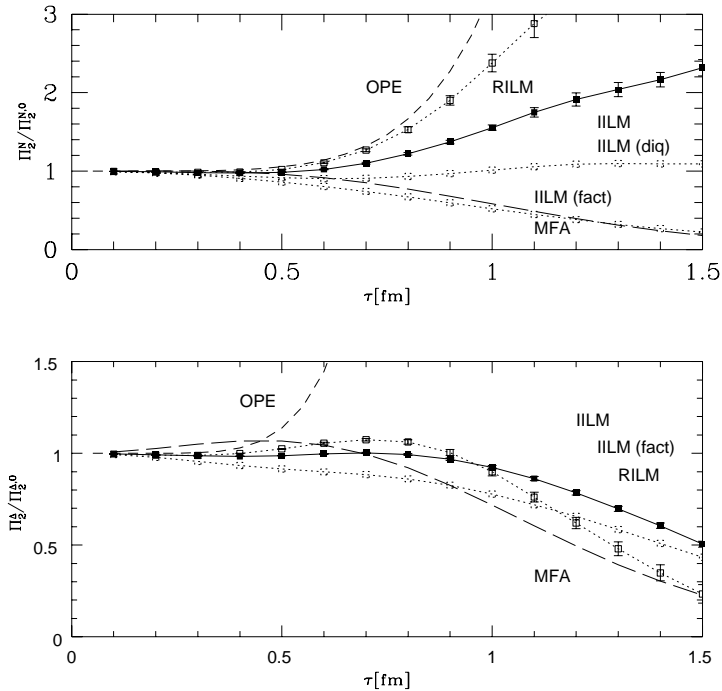}}
\vspace{-3mm}
\caption{\label{fig_bar_cor}
Nucleon and delta correlation functions $\Pi_2^N$ and $\Pi_2^\Delta$. Curves
labeled as in Figs.~\ref{fig_pi_cor}-\ref{fig_rho_cor}.}
\end{figure}

The correlation function $\Pi_2^N$ in the interacting ensemble is shown in 
Fig.~\ref{fig_bar_cor}. There is a significant enhancement over the perturbative
contribution which is nicely described in terms of the nucleon contribution. 
Numerically, we find\footnote{Note that this value corresponds to a relatively 
large current quark mass $m=30$ MeV.} $m_N=1.019$ GeV. In the random ensemble, 
we have measured the nucleon mass at smaller quark masses and found $m_N=0.96
\pm 0.03$ GeV. The nucleon mass is fairly insensitive to the instanton ensemble.
However, the strength of the correlation function depends on the instanton 
ensemble. This is reflected by the value of the nucleon coupling constant, 
which is smaller in the IILM.
In\cite{SSV_94} we studied all six nucleon correlation functions. We showed 
that all correlation functions can be described with the same nucleon mass and 
coupling constants.

The fitted value of the threshold is $E_0\simeq 1.8$ GeV, indicating that there
is little strength in the ``three quark continuum'' (dual to higher
resonances
in the nucleon channel). The fact that 
the nucleon in IILM is actually bound
 can also be demonstrated by comparing the full nucleon 
correlation function with that of three non-interacting quarks (the cube of 
the average propagator. The full correlator is 
significantly larger than the non-interacting one.

 Significant part of this interaction  was traced down to
strongly attractive {\em scalar diquarks}.
 The nucleon (at least in 
IILM) is a strongly bound diquark, plus a  loosely bound the third quark. 

The properties of this diquark picture of the nucleon continue to be
disputed
by phenomenologists. We would return to diquarks in the next section,
where
they would become Cooper pairs of Color Superconductors.

In the case of the $\Delta$ resonance, there exists only one independent 
current, given (for the $\Delta^{++}$) by
\be
\eta^\Delta_\mu =  \epsilon_{abc} (u^a C\gamma_\mu u^b)  u^c.  
\ee
However, the spin structure of the correlator $\Pi^\Delta_{\mu\nu;\alpha\beta}
(x)=\langle\eta^\Delta_{\mu\alpha}(0) \bar \eta^\Delta_{\nu\beta}(x)\rangle$
is much richer. In general, there are ten independent tensor structures, but 
the Rarita-Schwinger constraint $\gamma^\mu \eta_\mu^\Delta=0$ reduces this 
number to four. 

The mass of the delta resonance is too large in the random model, but closer to
experiment in the unquenched ensemble. Note that similar to the nucleon, part 
of this discrepancy is due to the value of the current mass. Nevertheless, the 
delta-nucleon mass splitting in the unquenched ensemble is $m_\Delta-m_N=409$ 
MeV,  larger but comparable to the experimental value 297 MeV.
It mostly comes from the absence of attractive scalar diquarks in this channel.

\section{The Phases of QCD}
Turning to finite temperature/density  QCD now, let me start with
emphasizing its main goals. Those are  not to use
once again a semi-classical or perturbative calculations,
 similar to what have been done before in vacuum. What we are looking for
 here
are {\em new phases} of QCD (and related theories), namely 
new self-consistent solutions which differs qualitatively from
what we have in the QCD vacuum. 

One such phase occurs at high enough temperature $T>T_c$: it is known
as
Quark Gluon Plasma (QGP). It is a phase  understandable in
terms of basic quark and gluon-like excitations \cite{Shu_78}, without
confinement
and with unbroken chiral symmetry in the massless limit\footnote{
It does not mean though, that it is a simple issue to understand even
the high-T limit of QCD, related to non-perturbative 3d dynamics.}.
 One of the main goals of heavy ion
program, especially at new Brookhaven dedicated facility RHIC, is to
study transitions to this phase.

 Another one, which gets much attention recently, is the direction
of finite density. Very robust Color Superconductivity was found to
be the case
here. Let me also mention
one more frontier $not$ to be discussed below,
 which has not yet attracted sufficient
attention: namely transition (or many transitions?) as the number of
light flavors $N_f$ grows. The minimal scenario includes transition
from the usual hadronic phase to one more unusual QCD phase,
the $conformal$ one, in which there are no 
particle-like excitations and correlators
   are power-like in the infrared. Even the position of the critical
   point is unknown. 

That
the main driving force of these studies is the
intellectual challenge it provides. In order to prove to you that the
problems related with mechanism of confinement and chiral
symmetry breaking are not simple, it is enough to remind
that a lot of people who
moved these days into quantum gravity/string theory did so
partly because of
frustration with the non-perturbative QCD.
However I still think that, with the help of all experimental data, real and
numerical, we have at hand for QCD we still have much  better chances here.

\subsection{The Phase diagram}

\begin{figure}[ht]
\vskip .05in
 \begin{minipage}[c]{3.in}
 \centering
 \includegraphics[width=2.in, angle=270]{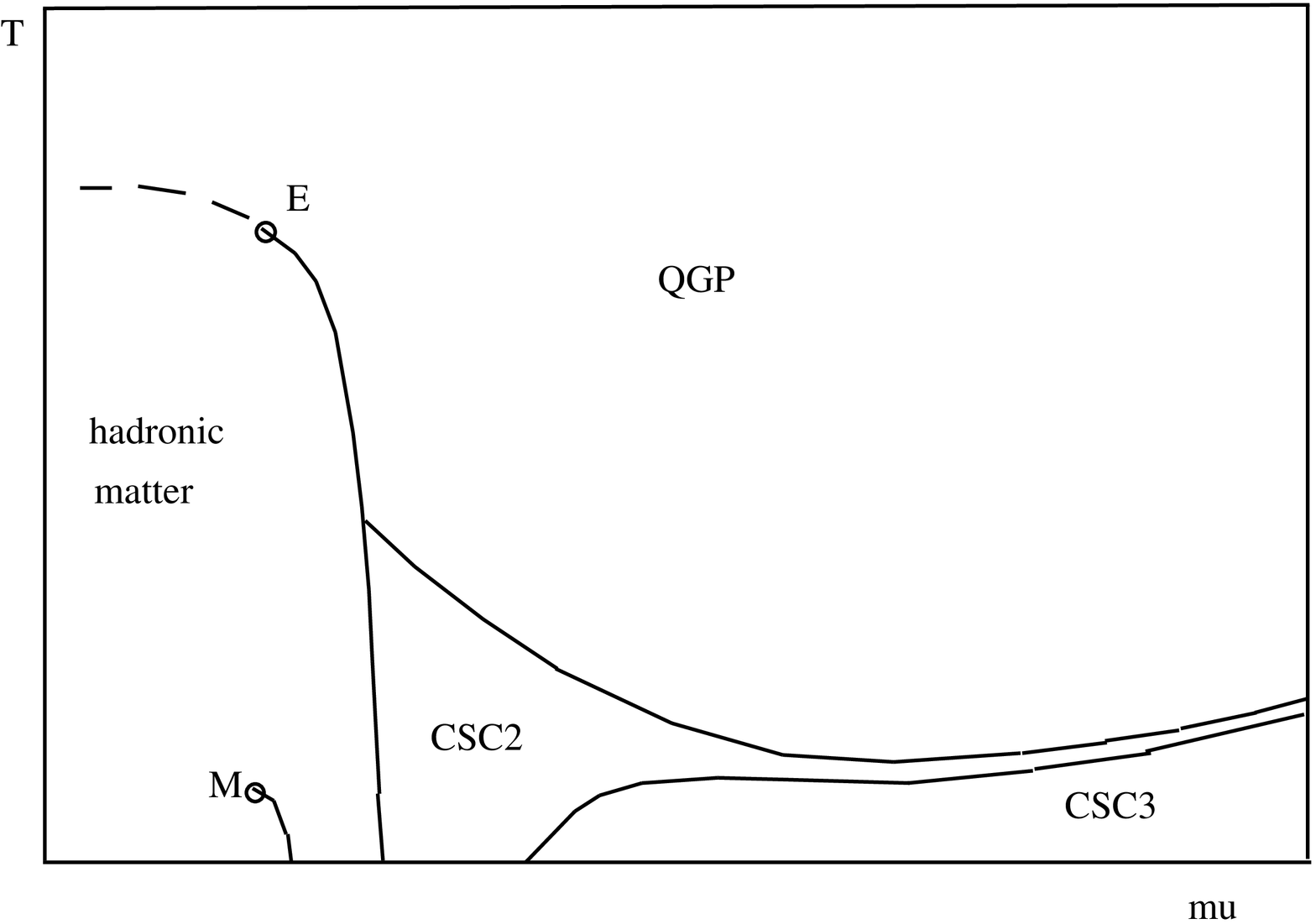}
 \end{minipage}
   \begin{minipage}[c]{3.in}
   \centering
\includegraphics[width=2.5in, angle=270]{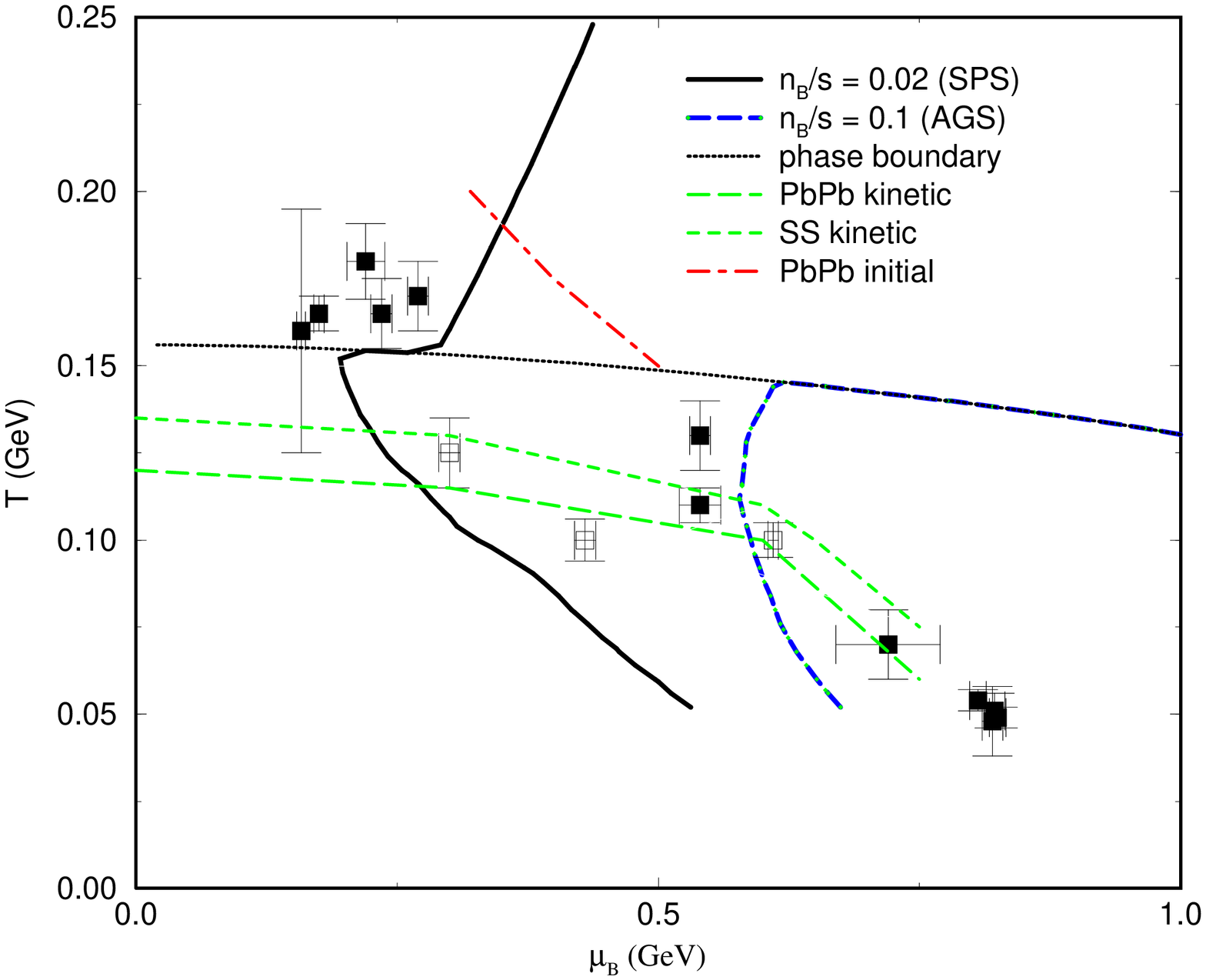}
 \end{minipage}
\vskip 0.1in
\caption{
\label{fig_phases_th}
(a) Schematic phase diagram of QCD, in temperature T- baryon chemical
potential $\mu$ plane. E and M show critical endpoints of first
order transitions: M (from multi-fragmentation) is that for liquid-gas
transition in
nuclear matter. The color superconducting phases, CSC2 and   CSC3 are
explained in the text.
  (b) The part of the phase diagram which has been  experimentally
  accessible
by heavy ion collisions. 
 Closed (open) points correspond to chemical
  (thermal) freeze-out.
The zigzag is the adiabatic
paths correspond to entropy per baryon ratio indicated.
 Lines of thermal freeze-out for central  sulfur-sulfur and PbPb
collisions,
as well as the
initial line for  PbPb are also indicated.
 }
\end{figure}

  The  QCD phase diagram version circa 1999  \cite{RSSV2}
is  shown in Fig
\ref{fig_phases_th}(a), at the baryonic chemical potential $\mu$
(normalized per quark, not per baryon) - the temperature T
plane. Some part of it is old: it is hadronic phase at small values
of both parameters, and QGP phase at large T,$\mu$.

The phase transition line separating them most probably does not really
start at $T=T_c,\mu=0$ but at a ``endpoint'' E, 
a remnant
  of the so called QCD tricritical point which QCD has in the chiral
(all quarks are massless) limit. 
Although we do not know where it is\footnote{Its position is very sensitive to
precise
value of the strange quark mass $m_s$}, we hope we know how to find it, see
\cite{SRS}. The proposed ideas rotate around the fact that the
 order parameter, the VEV of the sigma meson, is
at this point truly massless, and creates a kind of a ``critical opalesence''. 
Similar phenomena were predicted and then indeed observed at the
endpoint of another line (called M from multi-fragmentation), separating
liquid nuclear matter from nuclear gas phase. 

The 
large-density (and low-T) region
looks rather different from what was shown at
 conferences just few months ago: 
 there appear two new 
 Color Super-conducting phases there.  Unfortunately
  heavy ion collisions do not cross this part
of the phase diagrams, and so  it belongs to a neutron star
physics.

The 2-flavor-like color superconductor
CSC2 phase was known before \cite{earlysuper}, but realization that
it should  be induced by instantons \cite{RSSV,ARW} has increased 
the gaps (and $T_c$) from a few MeV
scale to $\sim$100 MeV (50 MeV). In the hindsight
 is should hardly be surprising, since the $same$ interaction in $\bar
q q$ channel is responsible for chiral symmetry breaking, producing the gap
(the constituent quark mass) as large as 350-400 MeV.
Furthermore, in the {\it two-color} QCD, there is the so called Pauli-Gursey
symmetry which relates these two condensates. So at high density the
chiral condensate $<\bar q q>$ simply rotates into superconductor one
 $<q q>$, while the gap remains the same.

The symmetries of the CSC2 phase are similar to
the 
electroweak part of the Standard Model, with the
condensed scalar isoscalar $ud$ diquark operating   as
Higgs. The colored condensate breaks the color group,
making 5 out of 8 gluons massive.
The 3-flavor-like phase, CSC3, is brand new: it was proposed in \cite{ARW2} based on
one gluon exchange interaction, but in fact it is favored by
instantons
as well \cite{RSSV2}. Its unusual features include {\it color-flavor
  locking}
and {\it coexistence} of both types of condensates, $ <qq>$ and $< \bar q q>$.
It combines features of the Higgs phase (8 massive
gluons) and of the usual hadronic phase (8 massless ``pions'').

   Above I mention approach to high density starting from the vacuum.
One can also work out in the opposite direction, starting from 
 very large densities and going down. As it was noticed in ref. \cite{Son},
electric part of one-gluon exchange is screened,
and therefore the Cooper pairs appears due to magnetic forces.
It  is
 interesting by itself, as a rare example: 
one has  to take care of {\em time delay effects} of the interaction.
 The result is the indefinitely growing gaps
  at  large $\mu> 10 GeV$, as 
$ \Delta \sim \mu exp( -{3\pi^2 \over \sqrt{2}g(\mu)})$.

\subsection{Finite T transition and Large Number of Flavors}
  There is no place here to discuss this subject in details:
there are rather extensive lattice data now, and so we actually
know quite a lot about these transition.

  Let me only emphasize what it is look like from the 
perspective of the instanton-based theory. If the near-random
set of instantons leads to chiral symmetry breaking and quasi-zero
modes at low T, we should be able to explain in the same terms
how the high-T phase look like. 
The simplest solution would be just disappearance of instantons
at $T>T_c$, and at some early time people thought this is what actually
happens. However, it should not be like this because the Debye
screening which is killing them only appears at $T=T_c$. Lattice
data works have also found no depletion of the instanton density
up to  $T=T_c$.

  On the other hand,
the absence of the condensate and quasi-zero modes
can only mean that the ``liquid" is now broken into finite pieces.
The simplest of them are pairs, or the instanton-anti-instanton
molecules. This is precisely what instanton simulations have found \cite{SS_98}.
Whether it is or is not so on the lattice is not yet clear. Some
nice molecules were located, but the evidences 
for the molecular mechanism are still far from being completely convincing.
No alternative have been so far  proposed, however.

  The results of simulations with $N_f=2,3,5$\footnote{The case 
$N_f=4$ is omitted because in this case it is very hard to determine 
whether the phase transition happens at $T>0$.} flavors with equal 
masses can be summarized as follows. 
 For $N_f=2$ there is second order phase transition 
which turns into a line of first order transitions in the $m-T$ plane
for $N_f>2$. If the system is in the chirally restored phase ($T>T_c$) 
at $m=0$, we find a discontinuity in the chiral order parameter if 
the mass is increased beyond some critical value. Qualitatively, the 
reason for this behavior is clear. While increasing the temperature 
increases the role of correlations caused by fermion determinant, 
increasing the quark mass has the opposite effect. We also observe 
that increasing the number of flavors lowers the transition temperature. 
Again, increasing the number of flavors means that the determinant
is raised to a higher power, so fermion induced correlations become
stronger. For $N_f=5$ we find that the transition temperature drops
to zero and the instanton liquid has a chirally symmetric ground state, 
provided the dynamical quark mass is less than some critical value. 
  Studying the instanton ensemble in more detail shows that in this 
case, all instantons are bound into molecules.

   Unfortunately, little is known about QCD with different numbers of
flavors from lattice simulations. There are data
by the Columbia group 
for $N_f=4$. The most important result is that chiral symmetry
breaking effects were found to be drastically smaller as compared 
to $N_f=0,2$. In particular, the mass splittings between chiral
partners such as $\pi-\sigma,\,\rho-a_1,\,N(\frac{1}{2}^+)-N(\frac{1}{2}^-)$, 
extrapolated to $m=0$ were found to be 4-5 times smaller. This 
agrees well with what was found in the interacting instanton model:
more work in this direction is certainly needed.

\subsection{Color superconductivity} My interest was initiated by finding 
\cite{SSV_94} that in the instanton liquid model 
even without $any$ quark matter,
the  {\em ud scalar diquarks} are very
deeply bound,  by amount comparable to
 the constituent quark mass. So, 
phenomenological manifestations \cite{diquarks}
 of such diquarks have in fact  deep dynamical roots: they follow
 from the same basic dynamics
as the  ``superconductivity'' of the QCD vacuum, the chiral
 ($\chi$-)symmetry breaking. These
spin-isospin-zero
   diquarks are related to pions, and should be quite 
  robust element of nucleon (octet baryons) structure\footnote{As opposed to $\Delta$
  (decuplet) baryons.} . 

Another argument for deeply bound diquarks comes from 
bi-color ($N_c=2$) theory: in it the scalar diquark is degenerate
with pions.  By
 continuity from $N_c=2$ to $3$, 
 a trace of it
 should exist in real QCD\footnote{
Instanton-induced interaction strength in diquark channel is
$1/(N_c-1)$ of that for $\bar q\gamma_5 q$ one. It is the same at
$N_c=2$, zero for large  $N_c$, and is   exactly in
 between for $N_c=3$.}.
  
Explicit calculations with
instanton-induced forces for  $N_f=2,N_c=3$ QCD
have been made in two simultaneous\footnote{Submitted to hep-ph on the
  same day.} papers
 \cite{RSSV,ARW}. Indeed, a  
  very robust
Cooper pairs and
  gaps
$\Delta\sim$ 100 MeV were found. From then on, the field is booming.

Instantons create  the following amusing {\em  triality}: 
there are three attractive channels 
 which compete: (i)
the
{instanton-induced} attraction in
$\bar q q $ channel
leading to  $\chi$-symmetry breaking.
(ii) the
{instanton-induced} attraction in
$ q q $ which leads to color superconductivity.
(iii)
the
{\em light-quark-induced} attraction of $\bar I I $, which
leads to pairing of instantons
into {``molecules''} and a Quark-Gluon Plasma (QGP) phase without 
$any$ condensates.

 {\em How the calculations are actually made?}. 
 Analytically, mostly in 
  the mean field approximation,  similar to the original BCS
  theory
in Gorkov formulation. 
Total thermodynamical potential consists of 
{\em ``kinetic energy''} of the quark Fermi gas, including 
 mass operators of two types (shown in figure below).
The  {\em ``potential energy''} in such approximation is the
interaction Lagrangian convoluted  with all possible condensates.
 For example, instanton-induced one with $N_f=3$ leads to  two types of
diagrams shown in Fig.4, with (a)
{$<\bar q q>^3$} and (b)  {$< q q>^2 <\bar q q>$}.
Then one minimizes the potential over all condensates and get {\em gap equations}:
algebra may be involved because  masses/condensates are
{\em  color-flavor matrices.} 
\begin{figure}[h]
\leavevmode
\epsfxsize=5.cm
 \begin{minipage}[c]{2.9in}
 \centering 
\includegraphics[width=.8in, angle=270]{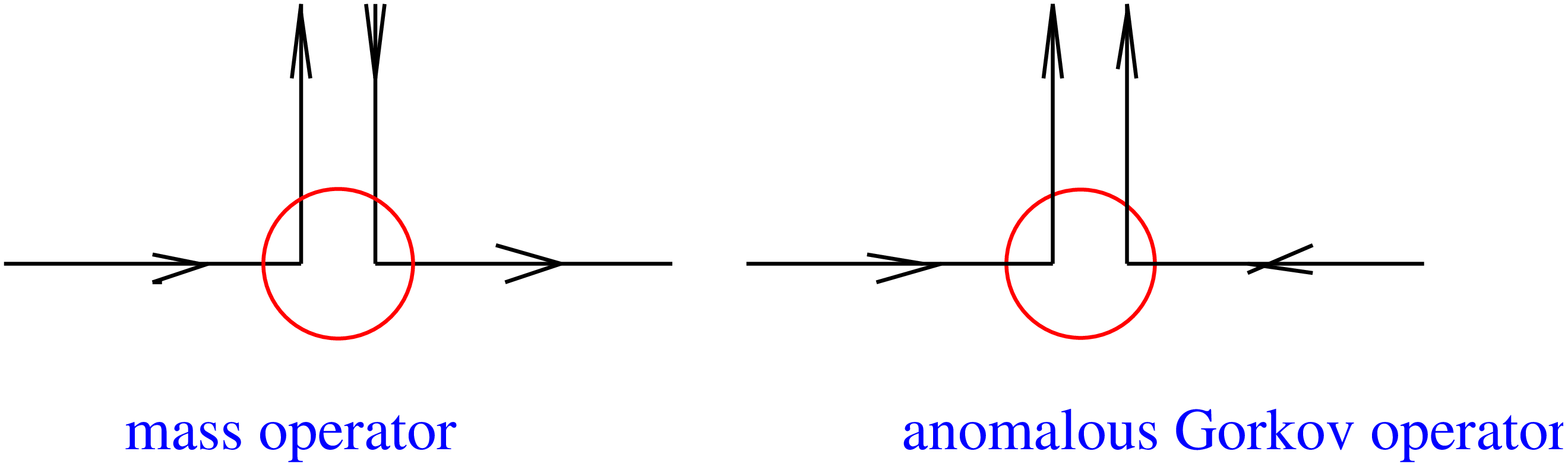}
\end{minipage}
 \begin{minipage}[c]{2.9in}
\centering
\includegraphics[width=1.in, angle=270]{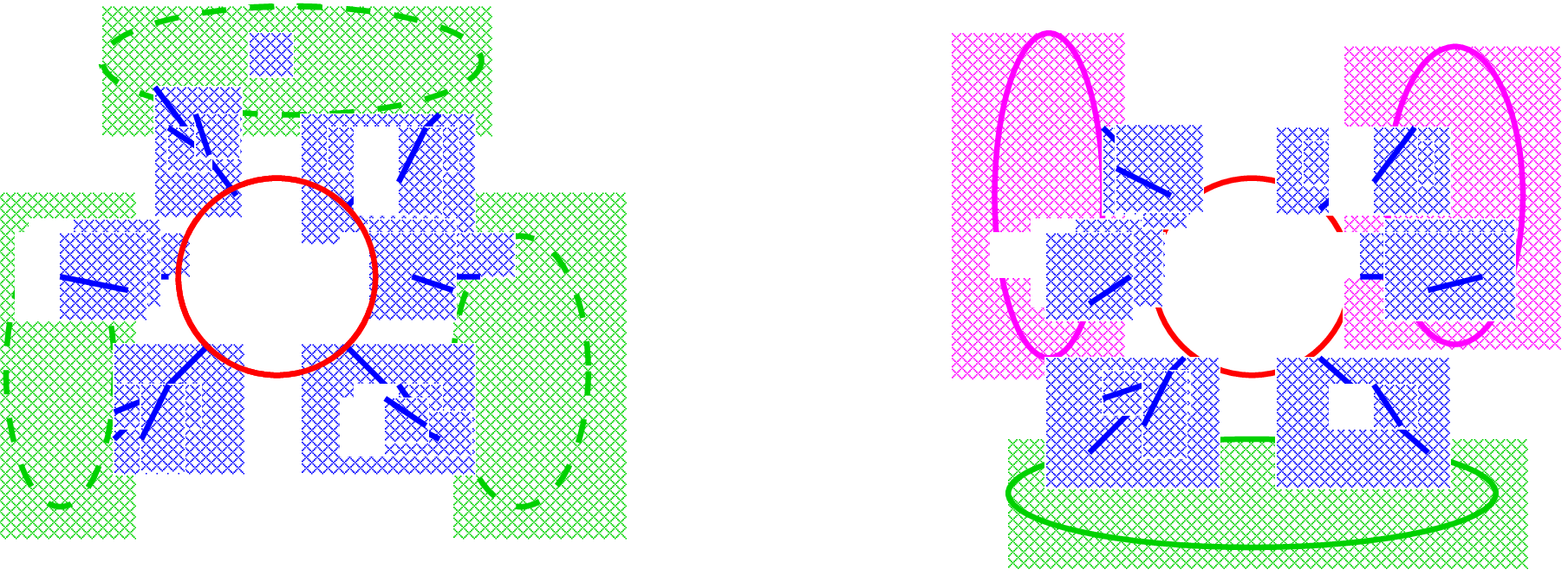}
\vskip -0.5in
\end{minipage}
\caption{(a) Two mass operators; (b) and example of two contribution
  to the potential energy for 3 flavor theory, including cube of the
$<\bar q q>$ and $|<qq>|^2<\bar q q>$.}
\end{figure}

%

{\bf Bi-color QCD: a very special theory  }
One reason it is special is well known to lattice community: its
fermionic determinant is $real$ even for non-zero $\mu$,
 which makes simulations possible.

 However the major interest to this theory
is related the so called {\em Pauli-Gursey symmetry}.
We have argued above that pions and diquarks appear at the same
one-instanton
level, and are so to say brothers.
In bi-color QCD they becomes identical twins:  due to additional
symmetry mentioned the diquarks are
{\em degenerate} with mesons.

In particular, the $\chi$-symmetry breaking is done like this
$SU(2N_f)\rightarrow Sp(2N_f)$, and
 for $N_f=2$ the coset 
$ K=SU(4)/Sp(4)=SO(6)/SO(5) =S^5$.
 Those 5 massless modes are
 pions plus scalar diquark S and its anti-particle $\bar S$. 

Vector diquarks are degenerates with vector mesons, etc. Therefore,
the scalar-vector splitting is in this case
about twice the constituent quark mass, or about 800 MeV. It should be
compared to binding in the ``real'' $N_c=3$ QCD of only 200-300 MeV,
and to zero binding in the large-$N_c$ limit.

The corresponding
sigma model describing this  $\chi$-symmetry breaking
was worked out in \cite{RSSV}: for further development 
see \cite{KST}.  As argued in  \cite{RSSV}, in this theory
the
critical value of transition to Color Superconductivity is simply
$\mu=m_\pi/2$, or zero in the chiral limit. The  diquark condensate is 
  just rotated $<\bar q q>$ one, and the gap is the constituent quark mass.
Recent lattice
works   \cite{2col} and
instanton liquid simulation \cite{S_dens} display it in great details, building
confidence for other cases.

{\bf Two flavor Three color QCD: the CSC2 phase}
 The first studies of instanton-induced CSC
were made for  this theory:  \cite{RSSV,ARW,CD}.
The gap around 100 MeV was found, and the main point is a {\em
  competition}
between the usual  $\chi$-symmetry breaking (pairing in the $\bar q q$
channel)
with CSC (pairing in the $ q q$
channel). In the corresponding high density phase we called CSC2
the chiral symmetry is simply restored.

In all these works one more  possible 
phase (intermediate between vacuum and CSC2), {\it Fermi gas of
  constituent quarks}, with
{both $M,\Delta\neq 0$} - was unstable. However in last more refined
calculation \cite{RSSV2} it obtains a small window.
 Its features are amusingly close
to those of nuclear matter: but it isn't, of course: to get nucleons
one should go outside the mean field. First attempted to do so in
\cite{RSSV2} was for  another cluster - the $\bar I I$
molecules. At T=0 it is however only 10\% correction to previous
results, but is dominant as T grows.

{\bf Three color Three flavor  QCD} produces a new phenomenon called
 {\em color-flavor locking} \cite{ARW2}. It  means that the
diquark condensate has the following  structure
$\langle q_i^a C q_j^b\rangle = \bar\Delta_1 \delta_{ia}\delta_{bj}
 + \bar\Delta_2 \delta_{ib}\delta_{ja}$, where ij are color and ab
 flavor indices. It is very symmetric,  reducing
$SU(3)_c SU(3)_f \rightarrow  SU(3)_{diagonal}$. We call it the CSC3
phase: it still breaks chiral symmetry.

 It was verified to be the lowest in \cite{ARW2} 
    for the one gluon exchange interaction
(adequate for high density): in this case only <qq>
condensate exists. About the same phase appears for the
instanton-induced Lagrangian (adequate for intermediate density) 
\cite{RSSV2}: but now the  $<\bar q q>$ condensates are non-zero as well.
The color-flavor locking is probably always the case for that theory.

Gaps $\delta_i$ and masses $\sigma_i$
(proportional to $<\bar q q>$), following from instanton-based
calculation \cite{RSSV2}, are shown as a function of $\mu$ in the
following figure 
\begin{figure}[h]
\vskip -1.cm
  \centering 
\includegraphics[width=3.6in]{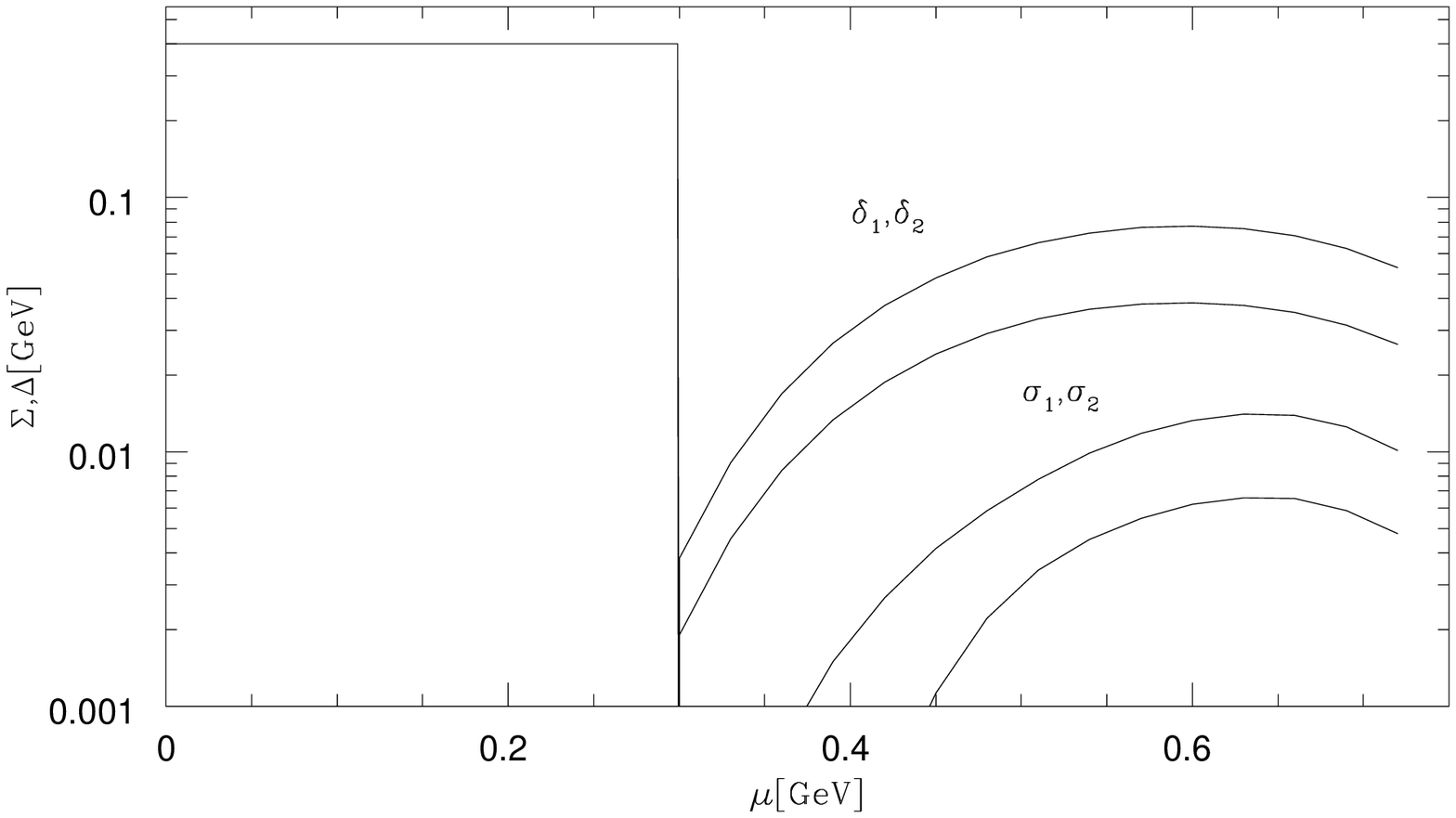}
\vskip -1.cm
\end{figure}

Physics issues under discussion for CSC3 include
{\em  hadron-quark continuity}. As 
 pointed out in \cite{SW_cont},  the CSC3 phase  
not only has the same  symmetries as hadronic 
matter (e.g. broken $\chi$-symmetry), but also very
similar
excitations. 
8 gluons become 8 {\em massive} vector mesons, 
3*3 quarks become 8+1  ``baryons''. The
8 massless pions  remain massless\footnote{Very exotic 3d objects,
  ``super-qualitons'' \cite{HRZ}, the skyrmions made of 
pions are among the excitations.}. Furthermore, photon and gluons
are combined into a {\em massless } $\gamma_{inside}$.
Calculation of masses and coupling constants of all of them is now in progress.
Can these phases be $distiquished$, and should there be $any$ phase transition
in the $N_f=3$ theory, separating it from nuclear matter? There is no 
need for it, at least from symmetry point of view.

{\bf The realistic  QCD}({Two} light plus {strange} flavor)
was studied in several papers \cite{RSSV2,2+1}.  Just
kinematically it is easy to see that  us,ds Cooper pairs with zero
momentum
is difficult to make: for $\mu_{u,d}=\mu_{s}$ the momenta
$p^F_{u,d}\neq p^F_{s}$.
Instantons generate also a dynamical operator $m_s (\bar u\bar d) (ud)$. 
Resulting behavior is as shown in our first figure.
 
{\bf Asymptotically large  densities }
   At $\mu >> 1 GeV$ the instantons are Debye-screened \cite{Shu_82}, as
   well as the electric (Coulomb) gluons. So
  {\em magnetic} gluons overtake electric ones   \cite{Son}.
 $Magnetically$ bound Cooper pair is
 interesting by itself, as a rare example: 
one has  to take care of {\em time delay effects} with
 Eliashberg eqn, etc. Angular integral leads to
second log in the gap equation, leading to   unusual answer:
$ \Delta \sim \mu \ exp( -3\pi^2 /\sqrt{2}g)$
which implies that \setcounter{footnote}{0}
 the gap {\em grows} indefinitely with $\mu$ \footnote{ 
Numerical details for all densities can be found in recent work
\cite{SW_high}.} and pQCD becomes finally justified. However, it is
the case for
huge densities, with $\mu >10 GeV$ or so.


{\bf Final remarks.}
All variants of CSC have an ``internal photon''
(a combination of the photon and gluon)  for which the
condensate is not charged and which therefore is not expelled
from it. So,  is it a superconductor, after all?

\setcounter{footnote}{0}
I think the answers is still ``yes''.
 For example, if one puts a  piece of CSC
into a magnet, it should still levitate: although
$\gamma_{inside}$
is massless,  the magnet uses $\gamma_{outside}$ field and a part of it
is expelled\footnote{The same would happen with
  a small piece of   Weinberg/Salam vacuum, if one can make
magnet with ``original''
(``outside'') field.}. 

Let me finish this section with few homework questions.
What is the role of confinement in all these transitions?
What is  nuclear matter for different quark masses, anyway?
 Do we have other phases in between, like diquark-quark phase
  or (analog of) K condensation, or different crystal-like phases?
 Is there indeed a (remnant of) the tricritical point which we can
find experimentally? And,  
  How can we  do finite density calculations on the lattice?

\section*{ Acknowledgments}
Let me thank the organizers of both schools, Chris Allton (in Swansea)
and Lidia Ferreira (in Lisbon),
for their kind  invitation and help.
This work is partially
supported by US DOE, by the grant No. DE-FG02-88ER40388.

\end{document}